\documentclass[10pt,preprint]{aastex} 

\newcommand{\citepeg}[1]{\citep[{e.g.,}][]{#1}}

\def\lsim{\hbox{ \rlap{\raise 0.425ex\hbox{$<$}}\lower 0.65ex\hbox{$\sim$}}}
\def\gsim{\hbox{ \rlap{\raise 0.425ex\hbox{$>$}}\lower 0.65ex\hbox{$\sim$}}}
\def\arcmin{\hbox{$^\prime$}}
\def\arcsec{\hbox{$^{\prime\prime}$}}
\def\arcdeg{\mbox{$^\circ$}}%

\def\fh{\hbox{$~\!\!^{\rm h}$}}
\def\fm{\hbox{$~\!\!^{\rm m}$}}
\def\fs{\hbox{$~\!\!^{\rm s}$}}
\def\ale{\mathrel{\hbox{\rlap{\hbox{\lower4pt\hbox{$\sim$}}}\hbox{$<$}}}}
\def\age{\mathrel{\hbox{\rlap{\hbox{\lower4pt\hbox{$\sim$}}}\hbox{$>$}}}}

\def\kms{km~s$^{-1}$ }

\begin{document}

\title{Closing in on a Short-Hard Burst Progenitor: Constraints from
Early-Time Optical Imaging and Spectroscopy of a Possible Host Galaxy
of GRB\,050509b}

\def\berk{1}
\def\lick{2}
\def\chandra{3}
\def\harvard{4}
\def\ias{5}
\def\kipac{6}
\def\psu{{7}}
\def\irvine{{8}}
\def\mit{9}
\def\davis{{10}}
\def\phys{{11}}
\def\carnegie{{12}}
\def\hubble{{13}}
\def\ssl{{14}}
\def\mich{{15}}
\def\lbl{{16}}
\def\jpl{{17}}

\author{J. S. Bloom\altaffilmark{\berk}, 
        J. X. Prochaska\altaffilmark{\lick},
        D. Pooley\altaffilmark{\berk,\chandra},
        C. H. Blake\altaffilmark{\harvard},
        R. J. Foley\altaffilmark{\berk},
        S. Jha\altaffilmark{\berk}, 
        E. Ramirez-Ruiz\altaffilmark{\ias,\lick,\chandra},
        J. Granot\altaffilmark{\kipac,\ias},
        A. V. Filippenko\altaffilmark{\berk}, 
        S. Sigurdsson\altaffilmark{\psu}, 
        A. J. Barth\altaffilmark{\irvine}, 
        H.-W. Chen\altaffilmark{\mit}, 
        M. C. Cooper\altaffilmark{\berk}, 
        E. E. Falco\altaffilmark{\harvard},
        R. R. Gal\altaffilmark{\davis},
        B. F. Gerke\altaffilmark{\phys},
        M. D. Gladders\altaffilmark{\carnegie}, 
        J. E. Greene\altaffilmark{\harvard}, 
        J. Hennanwi\altaffilmark{\berk,\hubble}, 
        L. C. Ho\altaffilmark{\carnegie}, 
        K. Hurley\altaffilmark{\ssl},
        B. P. Koester\altaffilmark{\mich},
        W. Li\altaffilmark{\berk},
        L. Lubin\altaffilmark{\davis}, 
        J. Newman\altaffilmark{\lbl,\hubble}, 
        D. A. Perley\altaffilmark{\berk},
        G. K. Squires\altaffilmark{\jpl}, and
        W. M. Wood-Vasey\altaffilmark{\harvard}}

\affil{$^\berk$ Department of Astronomy, 601 Campbell Hall, 
        University of California, Berkeley, CA 94720-3411.}

\affil{$^\lick$ University of California Observatories/Lick
Observatory, University of California, Santa Cruz, CA 95064.}

\affil{$^\chandra$ Chandra Fellow.}

\affil{$^\harvard$ Harvard-Smithsonian Center for Astrophysics, 60 Garden
Street, Cambridge, MA 02138.}

\affil{$^\ias$ Institute for Advanced Study, Olden Lane, Princeton, NJ 08540.}

\affil{$^\kipac$ KIPAC, Stanford University, P.O. Box 20450, Mail Stop
29, Stanford, CA 94309.}

\affil{$^\psu$ Department of Astronomy \& Astrophysics, 525 Davey Laboratory,
         Pennsylvania State University, University Park, PA 16802.}

\affil{$^\irvine$ Department of Physics \& Astronomy, 
  4129 Frederick Reines Hall, University of
      California, Irvine, CA 92697-4575.}

\affil{$^\mit$ Center for Space Research, Massachusetts Institute of Technology, Cambridge, MA 02139-4307}

\affil{$^\davis$ Department of Physics, One Shields Ave., University
        of California, Davis, CA 95616-8677.}

\affil{$^\phys$ Department of Physics, 366 LeConte Hall, University of
      California, Berkeley, CA 94720-7300.}

\affil{$^\carnegie$ Carnegie Observatories, 813 Santa Barbara Street,
            Pasadena, CA 91101.}

\affil{$^\hubble$ Hubble Fellow.}

\affil{$^\ssl$ UC Berkeley, Space Sciences Laboratory, 7 Gauss Way,
Berkeley, CA 94720-7450.}

\affil{$^\mich$ Department of Physics, University of Michigan, Ann
  Arbor, MI 48109-1090.}

\affil{$^\lbl$ Institute for Nuclear and Particle Astrophysics,
Lawrence Berkeley National Laboratory, Berkeley, CA 94720.}

\affil{$^\jpl$ Spitzer Science Center, California Institute of
Technology 314-6, Pasadena, CA 91125.}

%%%%%%%%%%%%%%%%%%%%%%%%%%%%%%%%%%%%%%%%%%%%%%%%%%%%%%%%%%%%%%%%%%%%%%%%%%
\begin{abstract}

The localization of the short-duration, hard-spectrum gamma-ray burst
\hbox{GRB}\,050509b by the {\it Swift} satellite was a watershed
event. Never before had a member of this mysterious subclass of
classic GRBs been rapidly and precisely positioned in a sky accessible
to the bevy of ground-based follow-up facilities. Thanks to the nearly
immediate relay of the GRB position by {\it Swift}, we began imaging
the GRB field 8 minutes after the burst and continued for the
following \hbox{8 days}.  Though the {\it Swift} X-ray Telescope (XRT)
discovered an X-ray afterglow of GRB\,050509b, the first ever of a
short-hard burst, no convincing optical/infrared candidate
afterglow or supernova was found for the object. We present a
re-analysis of the XRT afterglow and find an absolute position of
R.A. = 12\fh36\fm13.\fs59, Decl. = +28\arcdeg59\arcmin04\farcs9
(J2000), with a 1$\sigma$ uncertainty of 3\farcs68 in R.A., 3\farcs52
in Decl.; this is about $4$\arcsec\ to the west of the XRT position
reported previously. Close to this position is a bright elliptical
galaxy with redshift $z=0.2248 \pm 0.0002$, about $1'$ from the center
of a rich cluster of galaxies. This cluster has detectable diffuse
emission, with a temperature of $kT = 5.25^{+3.36}_{-1.68}$~keV.  We
also find several ($\sim$11) much fainter galaxies consistent with the
XRT position from deep Keck imaging and have obtained Gemini spectra
of several of these sources. Nevertheless we argue, based on
positional coincidences, that the GRB and the bright elliptical are
likely to be physically related. We thus have discovered evidence that
supports the notion that at least some short-duration, hard-spectra
GRBs are at cosmological distances.

We also explore the connection of the properties of the burst and the
afterglow, finding that GRB\,050509b was underluminous in both of
these relative to long-duration GRBs. However, we also demonstrate
that the ratio of the blast-wave energy to the $\gamma$-ray energy is
consistent with that of long-duration GRBs.  This suggests a
comparably high efficiency of $\gamma$-ray conversion as in long GRBs
as might be expected if the same emission mechanism is at work in
short and long GRBs.  Based on this analysis, on the location of the
GRB ($40 \pm 13$ kpc from a bright galaxy), on the galaxy type
(elliptical), and the lack of a coincident supernova, we suggest that
there is now observational consistency with the hypothesis that
short-hard bursts arise during the merger of a compact binary (two
neutron stars, or a neutron star and a black hole). In this context,
we limit the properties of a Li-Paczy\'nski ''mini-supernova'' that is
predicted to arise on $\sim$day timescales.  Other progenitor models
are still viable, and additional rapidly localized bursts from the
{\it Swift} mission will undoubtedly help to further clarify the
progenitor picture.
\end{abstract}


\section{Introduction}

The distribution in duration \citep{mgi+81,ncdt84} and hardness
\citep{kmf+93} reveals evidence for two distinct populations of
classic gamma-ray bursts (GRBs): long-duration bursts, with typical
durations around 30~s and peak energies at $\sim 200$ keV, and the
minority short-duration bursts, with durations of a few hundred
milliseconds (ms) and harder spectra.  Despite remarkable progress in
understanding the nature and progenitors of long-duration GRBs,
comparatively little has been learned about the origin of short-hard
bursts, primarily because very few such bursts have had rapid and
precise localizations.

The modeled bursting rate at redshift $z=0$ of long-soft bursts
outnumbers short-hard bursts by about a factor of 3.5 in the BATSE
catalog \citep{sch01}; this assumes the same bursting rate as a
function of redshift and does not include the effect of beaming,
which, if different for long and short bursts, would imply that the
intrinsic relative rates differ from those observed. While a number of
bursts have been triangulated through the Interplanetary Network (see
\citealt{hsk+05}) on roughly day-long timescales, there has only been
one precisely localized short-hard burst relayed to ground observers
in less than 1~hr (GRB\,050202/{\it Swift}:
\citealt{tab+05})\footnotemark\footnotetext{There have been a few
other short bursts (duration $\ale$2 sec) detected and well localized,
but with soft spectra and hence not members of the short-hard
class. For example, GRB\,040924 was a soft, X-ray rich GRB
\citep{fra+04,huf+05}. Hereafter, we use the term ``short burst''
interchangeably with short-hard burst.}; owing to its proximity to the
Sun at time of localization, sparse groundbased followup was
undertaken.  Including GRB\,050509b, this corresponds to a ratio of
1:18 for short-hard to long-soft burst detections with {\it Swift},
much smaller than the BATSE result.\footnotemark\footnotetext{As of 20
May 2005, {\it Swift} has localized 2 short bursts out of a total of
38; see
\url{http://swift.gsfc.nasa.gov/docs/swift/archive/grb\_table.html .}}

As with long-duration bursts, the distribution of short bursts appears
very nearly isotropic \citep{kmf+93,bpp+96}, and their brightness
distribution ($<V/V_{\rm max}> \approx 0.35$) is consistent with being
a cosmological population. Still, there is no strong evidence to
support the idea that short bursts are preferentially seen from $z
\ale 0.37$ rich Abell clusters \citep{hhk+97}, nor are they clearly
connected with star formation within $\sim$100 Mpc \citep{ngpf05}.

Without precise and rapid localizations, the population statistics do
not provide a strong constraint on the short-burst progenitors.
Still, it has been largely reckoned that the leading candidates for
short bursts are the merger of a neutron star binary (NS--NS;
\citealt{bnp84,pac86,pac91,npp92,kc96,rj99,ros02,rrd03}) or a black
hole--neutron star binary (BH--NS;
\citealt{ls76,elps89,mhim93,kl98,bethe99,pwf99,fwh99}).  These systems hold
several particular attractions. First, although uncertain, the
estimated rate of mergers (between 1.5--20 per 10$^{6}$ yr per galaxy;
\citealt{bkb02,ss02,rrd03}) is comparable to the short-burst rate
\citep{sch01}. Second, the dynamical timescale of such mergers is
several milliseconds and the sound-crossing times are of order ten
milliseconds, comparable to the shortest observed bursts
\citep{mcm05}. Third, compact merger systems are likely to contain
enough mass-energy in a transient torus to power short-burst fluences
as would be observed if at cosmological distances
\citep{rrd03,lrp04,ros05}.  The typical dynamical timescale in such
binaries immediately prior to coalescence (ms) is much shorter than
the observed burst duration, and so it requires the central engine to
evolve into a configuration that is stable, while retaining a
sufficient amount of energy to power the burst \citep{lrp04}.

Mergers of such compact remnants are by no means the only possible
channel to produce short bursts. Evaporating primordial black holes
may produce short ($< 100$ ms) GRBs \citep{cmo99}, though basic
energetics arguments suggest that it would be difficult to see such
sources from distances well beyond the Galaxy. The recent discovery of
a megaflare from SGR 1806$-$20 \citep{mgv+05,hbs+05,pbg+05,ttt+05} led
to plausible suggestions that a substantial fraction ($\approx$40\%)
of short bursts could be produced by extragalactic magnetars
\citep{hbs+05}.  However, positional \citep{pbg+05,ngpf05} and
spectral \citep{lgg05} arguments have led other workers to suggest
that at most a few percent of the BATSE catalog could consist of
short-burst magnetars.  Note that not all compact mergers create
fertile conditions (a transient torus around a BH) for making a short
burst \citepeg{jr01,rsw04}.  The duration of the burst in a compact
binary merger is determined by the viscous timescale of the accreting
gas, which is significantly longer than the dynamical timescale, thus
accounting naturally for the large difference between the durations of
bursts and their fast variability \citep{lrp04}. In the collapsar
scenario for long-duration bursts, on the other hand, the burst
duration is given by the fall-back time of the gas \citep{woo93,mw99},
which is typically greater than a few seconds. However, a modified
collapsar scenario in which the burst duration is determined not by
fall-back but rather by the dynamical timescales associated with the
expanding outflow might still meet the constraints of short GRBs
\citep{woo01}.

The theoretical predictions for the afterglows of short GRBs have been
considered by \citet{pkn01}. Since the peak flux of the prompt
emission is comparable for short and long GRBs, if their distance
scales are similar the isotropic equivalent energy output in gamma
rays ($E_{\rm\gamma,iso}$) would be proportional to the duration of
the GRB, which is $\sim 10-100$ times larger for long GRBs. If the
efficiency for producing the gamma rays is comparable, then the
isotropic equivalent kinetic energy in the afterglow shock ($E_{\rm
k,iso}$) would have a similar ratio between long and short GRBs. This
would imply the afterglow of short GRBs to be on average $\sim 10-40$
times dimmer in flux than that of long GRBs. The afterglows of short
GRBs would be even much dimmer than this if they encounter a much
smaller external density compared to long GRBs; this is the
expectation from short-bursts from binary mergers outside of the host
galaxy. \cite{pkn01} argued that a low external density would not
affect the X-ray band, as the latter was assumed to lie above the
cooling-break frequency, $\nu_c$. We find that for a very low external
density the electron cooling becomes very slow so that $\nu_c$ can lie
above the X-ray band for the first few days, thus reducing the X-ray
flux compared to that for a higher external density typical of the
interstellar medium (ISM) found near star-forming regions of
long-duration GRBs.

To date the deepest early-time observations ($\Delta t \ale 1$ hr)
yielded upper limits $V_{\rm lim} \approx 14$ mag from the 0.3~m
ROTSE-I experiment \citep{kab+01}. \citet{hbc+02} compiled deeper
non-detections at optical and radio wavelengths at times from days to
weeks after four short bursts, with the faintest non-detection of $R
\approx 22.3$ mag at $\Delta t = 20$ hr (see also
\citealt{gan+00}). Clearly, deep and early observations in search of a
short-burst afterglow would require a rapid localization to an
uncertainty comparable to the field of view of meter-class (and
larger) telescopes.

GRB\,050509b \citep{geh+05} triggered the BAT coded-mask imager
on-board {\it Swift} on 9 May 2005 04:00:19.23 (UT dates and times are
used throughout this paper; \citealt{hrb+05}). The position of
GRB\,050509b, with an uncertainty of $4$\arcmin\ radius, was relayed
to the ground within a few seconds. The initial localization was later
revised to a position R.A.\ = 12\fh36\fm18\fs, Decl.\ =
+28\arcdeg59\arcmin28\arcsec, with a 95\% confidence error radius of
2.8\arcmin\ \citep{bbc+05}. \citet{bbc+05} describe the burst as a
single-peaked source with duration of $\sim$30 ms, peak flux of 2100
counts s$^{-1}$ (15--350 keV), and a hardness ratio consistent with
that of the short-hard population. At 06:29:23, a fading X-ray source
was reported with a 6\arcsec\ localization \citep{kbn+05} and later
updated to an 8\arcsec\ uncertainty radius at position R.A. =
12\fh36\fm13.9\fs, Decl. = +28\arcdeg59\arcmin01\arcsec\
\citep{rpb+05}.

GRB\,050509b thus represents the first short-hard burst localized in
real time to a position suitable for immediate follow-up observations
from a suite of ground-based facilities.  In this paper we describe
the results of our observations of the field of GRB\,050509b and what
bearing these data have on the nature of short bursts and the physics
of short-burst afterglows. In \S \ref{sec:obs} we describe
imaging and spectroscopy of the field. Our analysis of the X-ray
afterglow of GRB\,050509b is given in \S \ref{sec:xray}, leading to a
localization near an elliptical galaxy (\S \ref{sec:association}).  In
\S \ref{sec:host} we present a spectrum of that galaxy, its redshift,
and inferred properties. We then argue, on statistical grounds, for a
plausible association of this galaxy and the GRB. We demonstrate in \S
\ref{sec:theory} how GRB\,050509b appears to be a subluminous burst
relative to long-duration GRBs, but with a ratio of blast-wave energy
to gamma-ray energy that is consistent with the long-duration
population. In the remaining sections we describe new constraints on
the nature of short-burst progenitors. Throughout, we assume a
concordance cosmology with $H_0 = 70$ \kms Mpc$^{-1}$, $\Omega_\Lambda
= 0.7$, and $\Omega_m = 0.3$. All of the results presented herein,
though generally consistent with our previous results in GCN
Circulars, supersede them.

\section{Observations and Reduction}
\label{sec:obs}

\begin{figure*}[p] 
\centerline{\includegraphics[width=3.2in,angle=0]{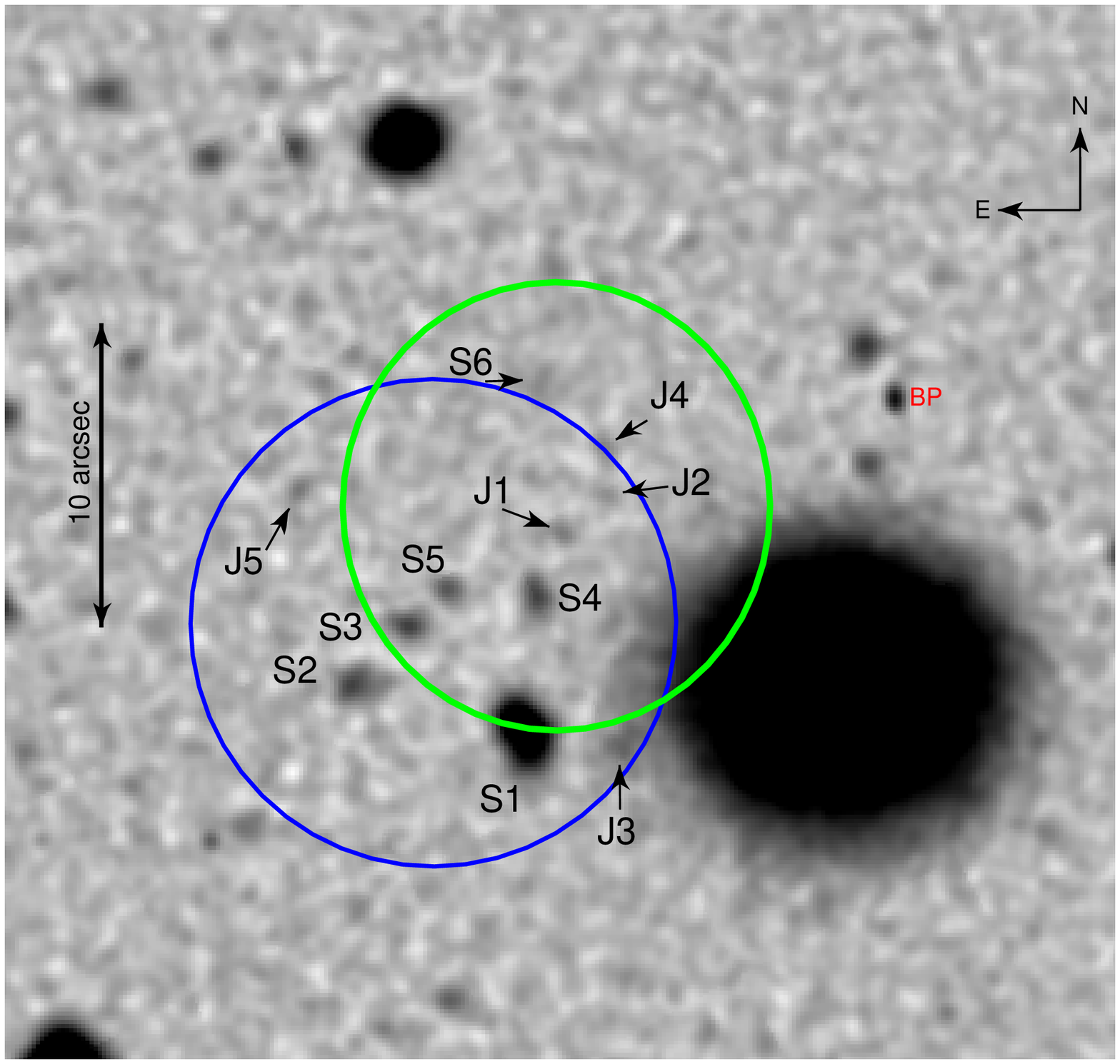}}
\centerline{\includegraphics[width=3.2in,angle=0]{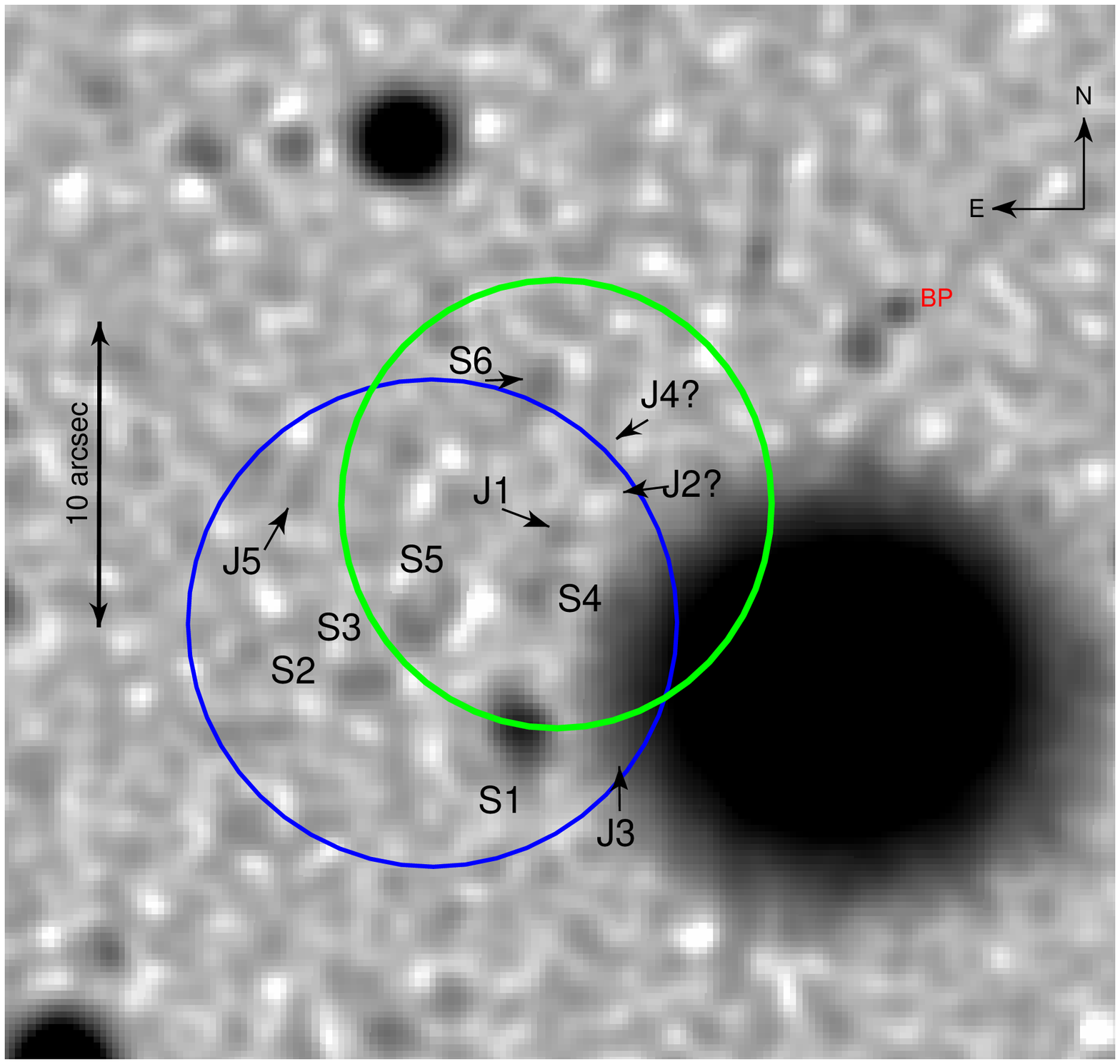}}
\caption[XRT Error Circle] {\small Keck/LRIS $G$-band (top) and
$R$-band (bottom) images, zoomed to show the XRT error circle. The
larger, blue circle is the revised XRT position from \citet{rpb+05};
the green circle to the west and north of that is our $2\sigma$
confidence region (see text) of the XRT position. The images have been
median subtracted and smoothed to accentuate the detection of faint
sources under the glare of the bright galaxy G1. The 11 sources
consistent with the \citet{rpb+05} X-ray afterglow localization are
labeled in both images, with the astrometric positions given in Table
\ref{tab:blobs}. North is up and east is to the left.  G1 is the large
galaxy to the west and south of the XRT. Bad pixel locations are denoted
with ``BP.''}
\label{fig:lris}
\end{figure*}

Initially, several groups reported \citep{rssq05,uce+05,bbp+05,tori05}
no new optical/infrared source that was consistent with the XRT
position of GRB\,050509b \citep{kbn+05}.  At 07:21:27 we
highlighted the proximity of the XRT to a bright red galaxy (hereafter
G1 = 2MASX J12361286+2858580) and suggested a plausible physical
association \citep{bbp+05} based on its presumed membership in a $z \approx 0.22$ cluster \citep{bbc+05}. We later reported the determination of the
redshift in \citet{pcn+05} and \citet{pbc05}. At 08:44:13 we
noted the presence of a faint, compact source (hereafter S1; see
\ref{fig:lris}) in the outskirts of G1, which we deemed a plausible
candidate counterpart \citep{bbp+05a}. A very similar suggestion was
made at 09:36:49 by \citet{csb+05}; in addition, they
noted apparent variability of the candidate (later retracting the
variability claim in \citealt{csb+05a}) and detection of three other
faint sources (S2--S4) consistent with the XRT position (see also
\citealt{csb+05b}). Two additional sources (S5 and S6) in the XRT
location were subsequently noted from Very Large Telescope (VLT)
imaging by \citet{hsj+05}, followed by another 5 sources (J1--J5)
reported by \citet{bfp+05}. No radio emission \citep{sf05,hww05} or GeV/TeV
emission \citep{park05} is consistent with the XRT error
localization. Below we discuss the observations, and further
interpretation, leading to these reports.

\subsection{Optical and Infrared Imaging}

We observed the field of GRB\,050509b on May 9 with the WIYN 3.5~m
telescope and the OPTIC CCD imager with a 9.6\arcmin $\times$
9.6\arcmin\ field of view and a plate scale of 0.14\arcsec/pixel.
Under poor ($\sim$2\arcsec) seeing conditions, two exposures totaling
360~s were obtained in the $i'$ band beginning at 04.344\,hr. In
addition, we obtained 2400~s of integration in the $r'$ band under
improved seeing conditions ($\sim$1\arcsec) beginning at 06.088\,hr.

The data were reduced in the usual manner using flat-fields from both
the illuminated dome and the twilight sky. The astrometric solutions
to the individual images were calculated by comparison to the
USNOB-1.0 catalog with a root-mean-square (rms) residual of
0.1\arcsec. The photometric zero-points of the images were calculated
by comparison to more than 50 stars in the Sloan Digital Sky Survey (SDSS)
photometry provided by \citet{ehp05}. The zero-points of the WIYN
images are uncertain at about the 3\% level. Limiting magnitudes were
estimated from the histogram of fluxes in 10$^4$ seeing-matched
apertures placed randomly within the field. The dispersion ($\sigma$)
of a Gaussian fitted to this distribution was used to estimate the
$5\sigma$ limiting flux in each image, which was converted to a
magnitude using the known zero-point.

The bright galaxy G1 to the west of the XRT position contaminates a
significant portion of the 8\arcsec\ radius XRT error circle. We used
{\it galfit} \citep{phir02} to fit a smooth S\'ersic profile to this
galaxy in order to remove most of the contaminant light prior to
examining the XRT error circle. A series of 1000 seeing-matched
apertures placed randomly within the XRT error circle identified no
new sources. The faint galaxy S1 was detected at the $>5\sigma$ level
in our deeper $r'$ images.

Near-infrared images were obtained with the 1.3~m PAIRITEL in the
$J$,$H$, and $K_s$ bands (see \citealt{bbs+05}). Observations consisted of
a 1130~s integration comprised of 7.8~s dithered exposures beginning
at 04.1375\,hr. These data were reduced by median-combining
sets of individual exposures within a moving 5-minute window. The
resulting median was used to subtract the bright sky from the
individual images. Finally, all of the individual images were combined
to make high-resolution mosaics using a modified version of {\it
drizzle} \citep{fh97}. Zero-points were determined 2MASS stars in the
field. Upper limits in the $J$,$H$,$K_s$ mosaics were estimated using
the same technique as for the WIYN data.  The WIYN and PAIRITEL upper
limits, as well as limits reported in the literature, are shown in
Figure~\ref{fig:upper}.

We later imaged the field of GRB\,050509b with the Keck~I~10~m
telescope and the LRIS-B instrument \citep{occ+95} using the
dichroic D560 (50\% transmission point at 5696~\AA) with $G$ and $R$
filters.  Starting at 11.25 May 2005, beginning near
astronomical twilight, we took 5 dithered images in each band for a
total of 1660~s and 1620~s exposures in $G$ and $R$, respectively. The
data were reduced in the usual manner and combined, weighted by
exposure times.

On 17 May 2005, 08:05.5, 8.17~d after the GRB, we obtained deep
$R$(Ellis) (which is similar to Harris $R$; \citealt{bmre03}) imaging
on the Echellete Spectrograph and Imager (ESI; \citealt{sbe+02}) on
the Keck~II 10~m telescope. In the presence of bright glare from the
Moon, we combined several reduced images for an effective exposure
time of 960~s. Since there is a negligibly small color term in
converting Harris $R$ to
$R_c$,\footnotemark\footnotetext{\url{http://www.ast.cam.ac.uk/$\sim$wfcsur/technical/photom/colours/
.}}
we found a zero-point relative to the LRIS $R$ image. There are no new
sources to $R_c \approx 25.0$ mag (5$\sigma$), nor significant
variations of the faint sources in the XRT error circle.

\section{The X-ray Emission}
\label{sec:xray}

The {\it Swift} XRT \citep{bhn+00} began observations of GRB\,050509b
on 2005 May 9 at 04:00:56, approximately 61~s after the BAT
trigger.  The observations consisted of eleven blocks, each about 2.5~ks 
in duration (except the first observation of 1.6~ks and the last
observation of 1.8~ks), spread over a period of $\sim$21 hr. The
XRT operated in a number of different modes throughout the
observations. The most common (32.3~ks of exposure) and most useful
mode for this object was the ``Photon Counting'' mode, which retains
the full imaging and spectroscopic resolution of the instrument.  The
images are 480 $\times$ 480 pixels, with a scale of 2\farcs36 per
pixel.  The XRT point-spread function is energy dependent, with a
half-power diameter of 18\arcsec\ at 1.5 keV.  The energy resolution
is also a function of energy, varying from about 50~eV at 0.1~keV to
about 190 eV at 10~keV.

The first Photon Counting observation began at 04:01:20
\citep{kbn+05} and lasted 1640~s.  As noted in \citet{kbn+05} and
\citet{rpb+05}, a faint X-ray source is detected in this first of the
eleven observations, but it faded quickly below the background.  We have
obtained the XRT data from the {\it Swift} archive, and have analyzed
them to determine the position of this X-ray afterglow candidate as
well as to examine its variability.  We briefly review the data
reduction, and then we discuss the localization of the afterglow
candidate and attempt to quantify the decay.

\subsection{{\it Swift} Data Reduction}
Using the Level 1 data from the {\it Swift} archive, we ran the
{\it xrtpipeline} script packaged with the HEAsoft 6.0 software supplied by
the NASA High Energy Astrophysics Science Archive Research
Center\footnotemark\footnotetext{\url{http://heasarc.gsfc.nasa.gov/ .}}.
We used the default grade selection (grades 0 to 12) and screening
parameters to produce a Level 2 event file re-calibrated according to
the most current (as of 2005 May 15) calibration files in the {\it Swift}
database\footnotemark\footnotetext{
\url{http://heasarc.gsfc.nasa.gov/docs/heasarc/caldb/swift/ .}}. To
produce images for source detection, we used the {\it xselect} software
(also part of HEAsoft 6.0), with a filter to include only counts in PI
channels 30--1000 (corresponding to photon energies of 0.3--10
keV). The PI channel to photon energy conversion was accomplished with
the redistribution file swxpc0to12\_20010101v007.rmf from the
calibration database.  The effective area of the XRT at the position
of the afterglow candidate was determined with the {\it xrtmkarf} tool,
using the correction for a point source.

\subsection{X-ray Afterglow Localization}

A number of factors make the localization of this X-ray afterglow
difficult. It is intrinsically faint and superposed on diffuse X-ray
emission from a galaxy cluster at $z=0.22$ \citep{gcl+03}.  The
initial source detection was performed with the wavelet-based routine
{\it wavdetect} \citep{fkrl02}, supplied with the CIAO 3.2 software
package, which in our experience is quite good at detecting faint
sources.  We chose parameters appropriate for detecting point sources
in this XRT observation; the pixel scales considered were a $\sqrt{2}$
series starting at 4 pixels (4, 5.657, 8, 11.314, 16), and the
significance threshold was set at $4 \times 10^{-6}$, corresponding to
a $\sim$1 false positive detection of a point source in the image.  We
detect 22 compact sources in the entire 32.3~ks data set.

To study the properties of the afterglow candidate, we extracted the
all the events within an area of radius 10 pixels around the nominal
{\it wavdetect} position.  In the first observation of 1.6~ks, there
are 14 counts in this region. When examining a plot of the cumulative
distribution versus time, we noticed that the majority of the counts
from this region occurred in the first 300~s.  We therefore further
investigated this brief interval.

\begin{figure*}[p]
\centerline{\includegraphics[width=6.0in,angle=00]{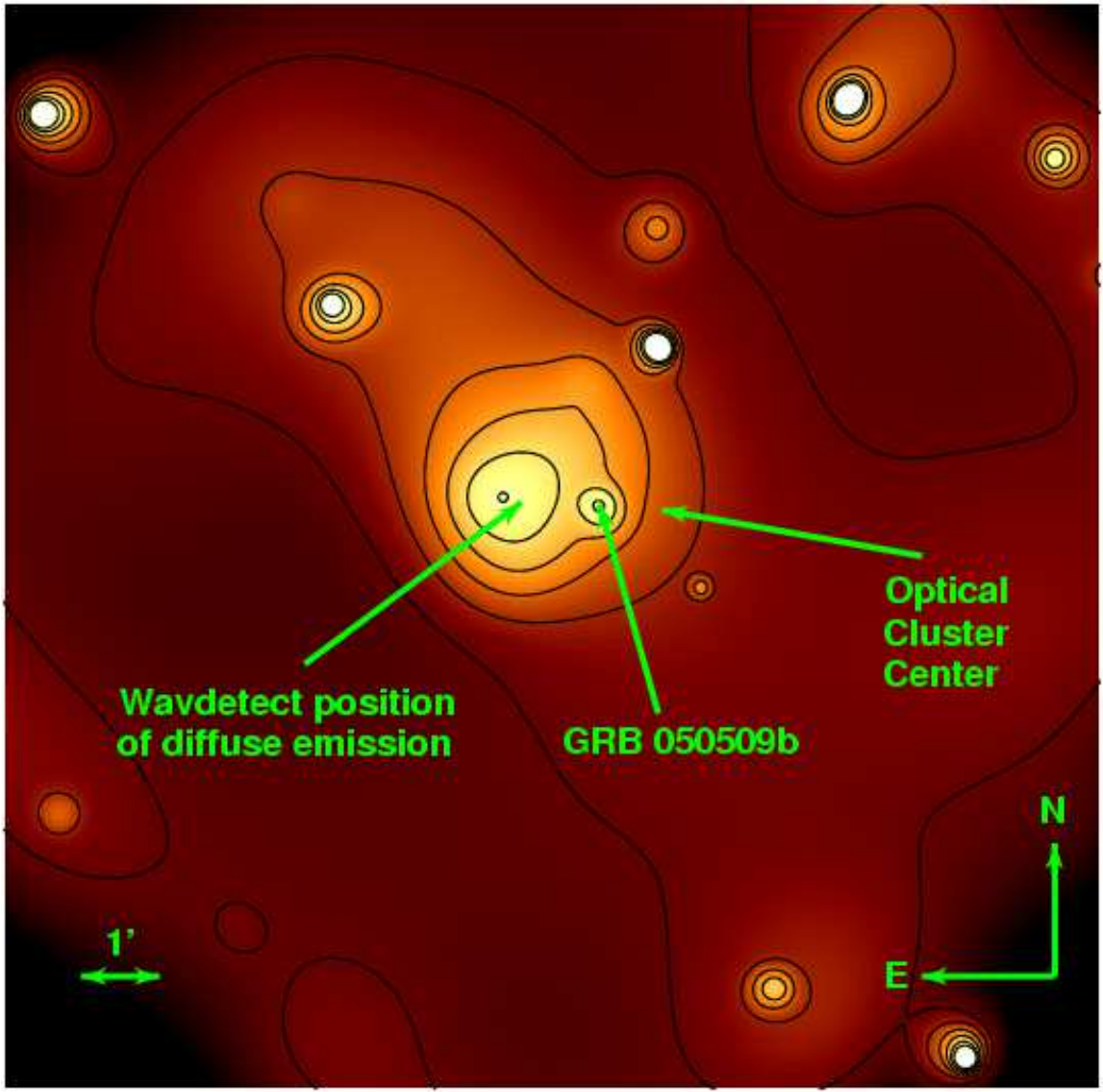}}
\caption{A $14\arcmin \times 14\arcmin$ adaptively smoothed image of
the XRT data.  GRB\,050509b is embedded in the diffuse X-ray emission
associated with the galaxy cluster, and is about 270~kpc in projection
from the cluster center.
\label{fig:xrt_smooth}}
\end{figure*}

In the first 300~s of the first Photon Counting observation, the XRT
detected 92 counts on the entire chip, with 73 of them outside of the
22 source regions.  Within any 10-pixel radius source region, we
therefore expect an average of 0.1 background counts.  We detect 9
counts in this region of the X-ray afterglow, with a reasonable
expectation that all 9 are from the X-ray afterglow.  Using the mean
location of just these 9 counts, we can obtain a relatively
uncontaminated estimate of the source position.  We calculate the 68\%
confidence interval in each direction as $T\sigma_j/\sqrt{N}$, where
$N$ is the number of counts (9), $\sigma_j$ is the sample standard
deviation of the 9 coordinates in each direction, and [$-T$,$T$] is
the 68\% confidence interval of the Student's $t$ distribution with
$N-1$ degrees of freedom. This gives us a position estimate, {\bf in
the {\it Swift} XRT reference frame}, of R.A.\ = 12\fh36\fm13\fs94,
Decl.\ = +28\arcdeg59\arcmin05\farcs3 (J2000) with an uncertainty of
3\farcs6 in R.A. and 3\farcs5 in Decl.  This is 4\farcs3 North of the
revised XRT position reported by \citet{rpb+05}.  A possible reason
for this offset is that the \citeauthor{rpb+05} position is based on
6.6~ks of XRT exposure and thus includes contributions from the
diffuse cluster emission (see Figure~\ref{fig:xrt_smooth}), biasing
the position estimate.

We examine the absolute astrometric accuracy of the {\it Swift} XRT
frame by searching for possible counterparts of the other 21 XRT
sources in deep optical images.  The best suited optical data for this
is a Bok $B$-band image \citep{ee05} because it covers an area large
enough to contain the entire XRT field.  Using a cross-correlation to
250 2MASS positions, we fit an absolute WCS using IRAF/CCMAP\footnote{IRAF is distributed by the National Optical Astronomy
Observatory, which is operated by the Association of Universities for
Research in Astronomy, Inc., under cooperative agreement with the
National Science Foundation.}. The overall geometry plus the
considerable distortion across the Bok $B$-band image was well fit by
a fourth-order polynomial with rms residuals of 0.135\arcsec\ in R.A.\
and 0.158\arcsec\ in Decl.  Assuming a 100~mas global uncertainty in
the 2MASS-International Celestial Reference System (ICRS) tie, the
absolute astrometry in the wide-field optical frame is thus uncertain
to 170 mas in R.A.\ and 187 mas in Decl.

\begin{figure*}[p] 
\centerline{\includegraphics[width=4.0in,angle=0]{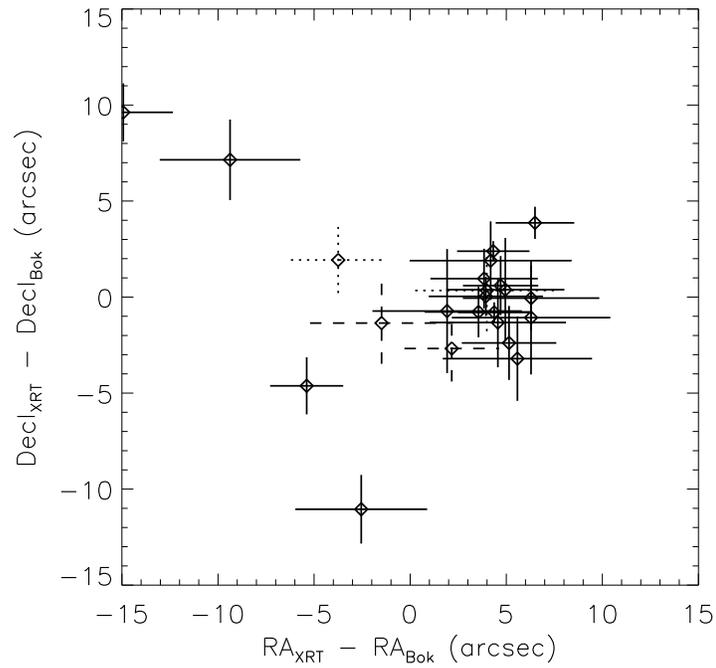}} 
\caption{Scatter plot of the position offsets between the XRT sources
(other than the afterglow candidate) and their nearest sources in the
wide-field Bok image B-band image. Two of the XRT sources each have
two Bok sources nearby. For these two XRT sources, the dashed line
represents the offset to the closest source, and the dotted line
represents the offset to the next closest source. There is strong
clustering around an offset of 4.5\arcsec\ in RA indicating a global
shift in the absolute astrometric zero-point of the XRT frame. The
number of X-ray/optical cross-matches is reasonable given the
sensitivity of the two frames (see text).}
\label{fig:x-opt-offset} \end{figure*}

For each XRT source other than the afterglow candidate,
Figure~\ref{fig:x-opt-offset} plots the offset between the XRT
position and the position of the closest optical source. Two XRT
sources had two optical sources within 5\arcsec; for these, the
closest optical source is represented by dashed lines and the next
closest by dotted lines.  There is an obvious locus around a
4\arcsec.5 difference in R.A., suggesting that these XRT sources are
associated with the corresponding nearest optical sources. At a
detection sensitivity around $10^{-14}$~erg~cm$^{-2}$~s$^{-1}$, it is
not surprising to find so many optical counterparts in the moderately
deep Bok image.  In a {\it Chandra}/Subaru study of the R.A. = 13~hr
{\it XMM}/{\it ROSAT} field, \citet{mgn+03} find unambiguous optical
counterparts for 61 of the 66 X-ray sources above
$10^{-14}$~erg~cm$^{-2}$~s$^{-1}$.  The mean $R$ magnitude of these
sources is $\bar{R}=20.7$, and the faintest counterpart is at $R=24.4$
mag.

Using the 14 sources in the above locus (excluding the two sources
with multiple possible counterparts), we derive an offset between the
XRT frame to the optical frame of 4\farcs49 $\pm$ 0\farcs72 W in
R.A. and 0\farcs42 $\pm$ 0\farcs30 S in Decl.  Our best estimate for
the location of the X-ray afterglow is therefore R.A. =
12\fh36\fm13\fs59, Decl. = +28\arcdeg59\arcmin04\farcs9 (J2000); this
is 4.1\arcsec\ west and 3.9\arcsec\ north of the revised XRT position
reported in \citet{rpb+05}.  The uncertainty in our position is a
combination of the statistical uncertainty of the XRT localization
(3\farcs6 in R.A., 3\farcs5 in Decl.) and the uncertainty in shifting
the XRT frame to the ICRS (0\farcs76 in R.A., 0\farcs40 in Decl.).

The astrometry in our original reports from WIYN and Keck imaging were
based on a frame of approximately 10 stars in the 2MASS catalog. The
release of the SDSS data and calibrations of this field allow us to
improve the astrometric tie to the ICRS. We fit the Keck/LRIS $G$-band
image to 91 sources in common with the SDSS object catalog with a
third-order polynomial solution using IRAF/CCMAP. The uncertainty in
the astrometric tie to SDSS, based upon residuals from the fit, is
$\sigma$(R.A.) = 0.134\arcsec and $\sigma$(Decl.) =
0.153\arcsec. Assuming a 75~mas astrometric uncertainty in the SDSS
astrometric calibration to the ICRS \citep{pmg+03}, we estimate the
absolute uncertainty in the Keck-ICRS tie is $\sigma$(R.A.) =
0.154\arcsec and $\sigma$(Decl.) = 0.171\arcsec.

The XRT location is $11.2$\arcsec $\pm$ 3.6\arcsec\ (or $40 \pm 13$
kpc in projection) from G1 as we first noted in
\citet{bbp+05}. Spectroscopy of this source reveals that it is indeed
an early-type galaxy (see \S~\ref{sec:host}) and is a member of a
cluster NSC J123610+285901 at $z \approx 0.22$ \citep{gcl+03,bbc+05}.
Near the location of the revised XRT error circle, we find $\sim$11
faint sources (all of which we or others have reported previously; see
above).  Figure \ref{fig:lris} shows the Keck $G$ and $R$ images with
identified source labeled. Table \ref{tab:blobs} gives the astrometric
positions and magnitudes of the sources.

\subsection{X-ray Afterglow Decay} 

We examine the first 1.6~ks block of observations to characterize the
temporal properties of the X-ray afterglow. A Kolmogorov-Smirnoff
(K-S) test on the arrival times of the 14 photons gives a probability
of 0.06\% that they come from a source with constant count rate.  The
next step in model complexity is one in which the X-ray count rate
$R_X$ in this region has a constant component (due to the background
and diffuse cluster emission) plus a component with a power-law
dependence on time (due to the fading afterglow). Our model is thus
$R_X(t) = A (t-t_0)^{-\alpha} + B$, where $B$ is the constant
(background plus cluster) count rate, $t_0$ is the time of the BAT
trigger, and $A$ is a normalization chosen such that the model
preserves the detected flux over the 1.6~ks under consideration.  We
determine $B$ to be 0.00107~count~s$^{-1}$ from the later
observations.

We considered a range of $\alpha$ from 0 to 4 and computed the K-S
probability of the observed data coming from the model for each value
of $\alpha$. The K-S probability was highest (97.8\%) at $\alpha
\approx 1.3$. For $\alpha \lesssim 1$ (0.77) and $\alpha \gtrsim 1.7$
(2.1), the K-S probability dropped below 32\% (5\%).  For
$\alpha=1.3$, the normalization $A$ is 22~count~s$^{-1}$.  We can
translate this to an energy-flux normalization by determining the
conversion from counts to erg~cm$^{-2}$.  We consider only the first
300~s of data for this determination in order to reduce contamination
from the background. For each of the 9 counts, we know its energy as
well as the effective area of the XRT at that energy.  The average is
$3.14 \times 10^{-11}$~erg~cm$^{-2}$~count$^{-1}$. Our model for the
X-ray flux (0.3--10 keV) of the afterglow only is then $F_X(t) = A'
(t-t_0)^{-\alpha}$. For $\alpha=1.3$, the normalization is $A'=6.9
\times 10^{-10}$~erg~cm$^{-2}$~s$^{-1}$. For example, the X-ray flux
of the afterglow at $t=200$~s after the BAT trigger for $\alpha=1$,
1.3, and 1.7 is $F_X(200) =$ 5.8, 7.0, and $6.5
\times10^{-13}$~erg~cm$^{-2}$~s$^{-1}$, respectively.

\begin{figure*}[p]
\centerline{\includegraphics[width=5in,angle=0]{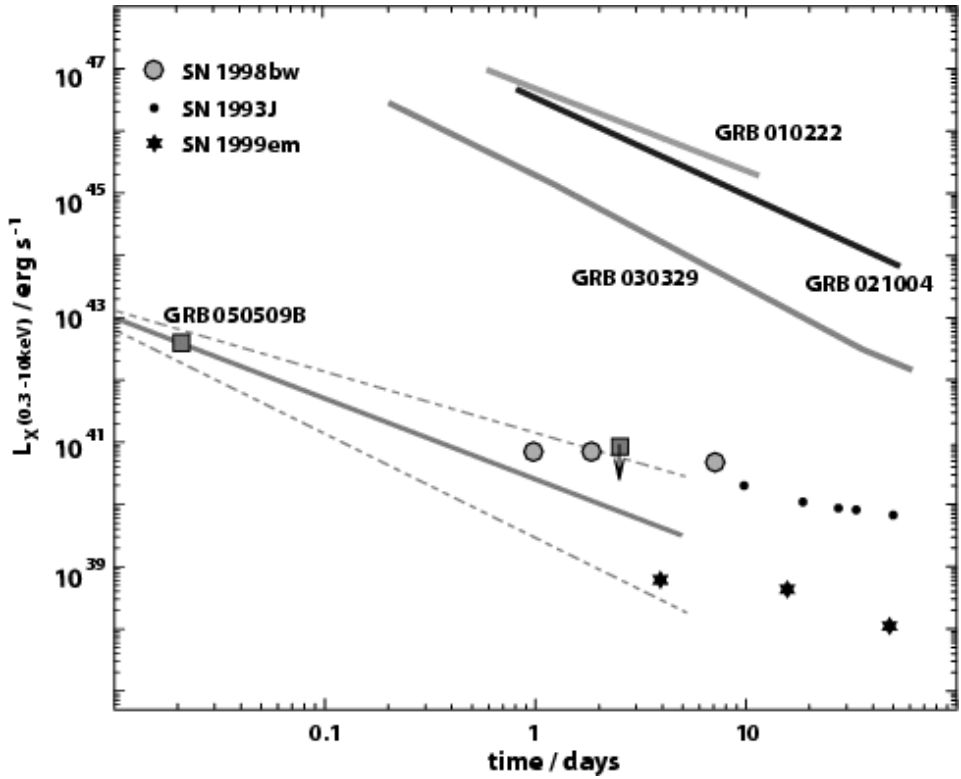}}
\caption{{\small Compilation of observed X-ray light curves for GRBs
and core-collapse supernovae (adapted from \citealt{kwp+04}). The X-ray
luminosities for the short GRB\,050509b are calculated assuming
$z=0.2248$.  For GRB 050509b we used data points (square symbols) from
both the XRT and the Chandra upper limit reported by \cite{patel05}.
The solid line corresponds to the best-fit temporal decay index of
$-1.3$.}}
\label{fig:xray}
\end{figure*}

Figure \ref{fig:xray} shows, on a common scale, X-ray light curves for a
number of GRBs and core-collapse supernovae. The location of the X-ray
transient associated with GRB\,050509b, when placed at 
$z=0.2248$, is striking. While the slope of the transient agrees well
with those of typical GRBs, its flux falls well below the typical 
long-GRB range. In fact, for any reasonable redshift (i.e., $z \lesssim
3$--5), GRB\,050509b would still be significantly underluminous in its
X-ray afterglow when compared to those of long-duration GRBs (see
Figure \ref{fig:xray}). For the assumed redshift of the tentative host
galaxy, the extrapolated X-ray luminosity at a few days, which is also
consistent with the {\it Chandra} upper limit \citep{patel05}, 
is close to those
seen in typical core-collapse supernovae.

\subsection{Diffuse Galaxy Cluster Emission} We used
{\it wavdetect} to search for large-scale structures in the full 
32.3~ks XRT data set.  The pixel scales searched were (20, 28.28, 40,
56.57, 80). The center of the diffuse emission presumably associated
with the galaxy cluster had a {\it wavdetect}-determined position of
12\fh36\fm18\fs26, +28\arcdeg59\arcmin06\farcs7.
Figure~\ref{fig:xrt_smooth} shows an adaptively smoothed image (using
the CIAO tool {\it csmooth}) of the XRT data with the cluster center
and GRB indicated.  The colors represent the 0.3--10 keV count
density. Contours are drawn at 0.00449, 0.00646, 0.00934, 0.0136,
0.0197, and 0.0273 count arcsec$^{-2}$.  As the image shows, the {\it
wavdetect}-determined position of the diffuse emission is about
14\arcsec\ to the west and 4\arcsec\ south of the peak of the diffuse
emission, which is at 12\fh36\fm19\fs33, +28\arcdeg59\arcmin10\farcs8
(J2000). [Note that the optical cluster center is 12\fh36\fm10
+28\arcdeg59\arcmin00.9\arcsec\ (J2000) as defined by the center of
the galaxy overdensity; this is about 125\arcsec\ east and 10\arcsec\
south of the peak of the diffuse X-ray emission.] We thus
find that the XRT afterglow position is 75\arcsec\ west, 6\arcsec\
north of the cluster center, as defined by the peak of the diffuse
X-ray emission, about 270~kpc in projection.

We extract a spectrum from a region of 110\arcsec\ in radius centered
on the {\it wavdetect} position.  We use a similar-sized region in a
source-free area to extract a spectrum for background subtraction.  We
require the cluster spectrum to contain at least 20 counts per bin,
and we consider the range 0.3--10~keV.  We fit the
background-subtracted cluster spectrum in Xspec v12.2 \citep{arna96}
with a MEKAL (warm plasma) model absorbed by a Galactic column density
of $1.52\times10^{20}$~cm$^{-2}$ \citep{dl90}.  We set the MEKAL
redshift at $z=0.2248$ and the metallicity at [Fe/H] = 0.26
\citep{ml97} and allow the temperature and normalization to vary.  The
best fit temperature is $kT = 5.25^{+3.36}_{-1.68}$~keV, which gives
$\chi^2/\mathrm{dof}=22.4/20$.

\section{Associating GRB\,050509b with G1}
\label{sec:association}

We are now in a position to explore the possible association of
GRB\,050509b with the cluster and with the nearby elliptical galaxy
G1. Focusing on the BAT localization alone, we first consider the
probability that a random position in the sky would be in a rich
cluster of galaxies (here we neglect the effects of lensing, expected
to be small; for example, \citealt{gn94}).  A reasonable estimate of
the covering fraction of clusters on the sky is given by the the DPOSS
Northern Sky Optical Cluster Survey \citep{gcl+03}.  Although this
survey is not very deep ($z_{\rm lim} \sim 0.3$), low-redshift
clusters should dominate the sky density.  \cite{gcl+03} find a
covering fraction of $\sim 0.03$ assuming a typical cluster radius of
1~Mpc, which suggests a chance alignment is improbable but not
impossible.  Moreover, the XRT localization of GRB\,050509b to within
45\arcsec\ of the center of such a cluster would occur by chance with a
probability of just $\sim 7 \times 10^{-4}$, but it is difficult to
estimate {\it a posteriori} how large a distance from a cluster center
one would have considered ``significant.''

While the gas from the cluster environment may enhance the probability
of localizing short-burst afterglows (see below), our expectation is
that short GRB progenitors are caused by the death of stars of some
sort, with the burst rate determined by processes on scales
significantly smaller than cluster lengths.  To this end, we should
consider the chance probability of the GRB event occurring at close
impact parameter to a galaxy similar to G1.  As reported by
\citet{ehp05}, the galaxy G1 has a Petrosian $r'$ magnitude of $17.18
\pm 0.02$ mag based on imaging by the SDSS.  The sky density of
galaxies with comparable apparent magnitude brighter than G1 is
$\sim 40$ per square degree \citep{blanton03}.  Therefore, the
probability of an event randomly occurring within 20\arcsec\ (about
twice the observed offset) of this bright galaxy is $\sim 5 \times
10^{-3}$. We consider this a conservative estimate because this
probability makes no reference to the galaxy redshift, type, size, or
age (which are consistent with {\it a priori} discussions of
short-burst host galaxies; e.g., \citealt{bpzy98} and
\citealt{bsp99}).

If one argues that GRB\,050509b is indeed physically associated with
this bright, low-redshift elliptical galaxy, one must consider why the
several other well-localized short bursts have not shown similar
associations. The first possibility, that the short bursts arise from
a more local population (as suggested by the magnetar flare from 27
December 2004; \citealt{hbs+05}) and GRB\,050509b must therefore
arise from a different population, was discounted for four of the
best-localized short bursts \citep{ngpf05}. Another possibility is
that GRB\,050509b was significantly closer than the other
well-localized short bursts. A strong test of this hypothesis is to
determine if other short bursts are associated with more distant
clusters or intrinsically bright, massive galaxies (e.g., through a
deep imaging campaign).  The third possibility is simply that
short-burst progenitors need not always arise in such galaxies. In
fact, for the NS--NS hypothesis we would expect mergers in galaxies
spanning a wide range of Hubble types.  A delayed BH--NS merger is also
possible, but less likely if GRB\,050509b is associated with G1:
statistically, the distribution of timescales for BH--NS coalescence
is as broad as that for NS--NS coalescence, albeit quite model
dependent \citep{bkb02,ss02}, but the systemic kick velocity is
expected to be systematically lower by a factor of a few in most
theoretical models of formation of BH--NS binaries (kick velocity is
roughly inversely proportional to mass, so more massive binaries
receive less kick). Moreover, larger velocity kicks generally lead to
shorter merger times.  Thus the $\sim$40 kpc offset tends to favor
NS--NS over BH--NS mergers.

While these possible associations are tantalizing, {\it a posteriori}
statistics are very suspect.  Had the GRB been near a bright spiral
galaxy, we might have made similar claims based on chance
probabilities.  Nevertheless, it remains the case that many workers
had predicted the distinct possibility that a well-localized compact
merger and/or short burst could be near an elliptical galaxy
\citep{bpzy98,bsp99,plp99} (see also \citealt{lsp+98}), and so we
suggest that these arguments might reasonably reflect a true
association. Moreover, the possible association with an early-type
host stands in stark contrast to results from long-duration GRBs
\citep{bkd02,ldm+03,dfk+03}, and is reminiscent \citepeg{bkd02,dar04}
of the dichotomy between core-collapse and thermonuclear (Type Ia)
supernovae. Only a larger sample of short GRBs will provide truly
compelling evidence for such a parallel. Still, based on the arguments
above, we proceed by accepting the hypothesis that GRB\,050509b is
physically associated with G1.

\section{A Putative Host Galaxy at $z$=0.2248}
\label{sec:host}

We obtained a spectrum of G1 with DEIMOS \citep{fpk+03} on the Keck~II
10~m telescope under photometric conditions.  The data were acquired
in a series of two exposures starting at 07.47\,hr on the
night of the burst.  The instrumental setup included the 600 line
mm$^{-1}$ grating blazed at 7500~\AA\ and centered at 7200~\AA, the
GG455 order-blocking filter, and standard CCD binning.  This setup gives
nearly continuous wavelength coverage in the range 4500--9000~\AA. We
observed the galaxy though a 1.1\arcsec\ $\times$ 20\arcsec\ slit at
sky position angle 90$^\circ$ and an airmass of 1.0.  This setup yields a 
FWHM resolution of $\sim 5$~\AA\ (i.e., $\sigma = 100$ km s$^{-1}$).  The
data were reduced and calibrated with the DEEP spectroscopic pipeline
for DEIMOS data \citep{cooper06}.  Wavelength calibration and
flat-fielding were performed using spectra of Xe-Ne-Kr-Ar and quartz
lamps (respectively) obtained that night.

The software provides a two-dimensional, sky-subtracted image of the
spectrum across two CCDs of the DEIMOS mosaic.  Unfortunately, the CCD
that includes the bluest data has a pair of blocked columns which lie
near the center of the galaxy profile.  Therefore, we extracted the 
one-dimensional (1D) spectrum on this CCD using optimal extraction techniques 
assuming a Gaussian profile with $\sigma = 9.2$ pixels (i.e., 1.1\arcsec).  
For the other CCD, we extracted a 1D spectrum by adopting a 26-pixel
(3\arcsec) boxcar aperture.  Finally, we processed and calibrated a
spectrophotometric standard star (BD+28\arcdeg4211) observed at the end of
this night.  After comparing its observed flux (in digital numbers)
against the STIS CALSPEC calibration\footnotemark\footnotetext{{\tt
ftp://ftp.stsci.edu/cdbs/cdbs2/calspec/}}, we calculated a sensitivity
function which could be applied to our galaxy spectra.

Spectroscopic observations of G1, S1, S2, and 2 unidentified sources
were obtained with the GMOS spectrometer \citep{hja+04} on the Gemini
North 8~m telescope beginning at 10.27 May 2005 under photometric
conditions.  We used a 0.75\arcsec\ slit, a R400 grating blazed at
7640~\AA, GG455 order-blocking filter, and set the central wavelength to
6500~\AA. The airmass was low (1.0--1.1), so the effects of
atmospheric dispersion were negligible \citep{fil82}. Standard CCD
processing and spectrum extraction were accomplished with IRAF using a
1.74\arcsec\ aperture for these sources of interest (S1,S2).  The data
were extracted using the optimal algorithm of \citet{horn86}.
Low-order polynomial fits to calibration-lamp spectra were used to
establish the wavelength scale.  Small adjustments derived from
night-sky lines in the object frames were applied.  Using techniques
discussed in \citet{wh88} and \citet{mfb+00}, we employed IRAF and our
own IDL routines to flux-calibrate the data and to remove telluric
lines using the well-exposed continua of the spectrophotometric
standard EG--131 \citep{bess99}.

\begin{figure*}[p] 
\centerline{\includegraphics[width=6.5in,angle=90]{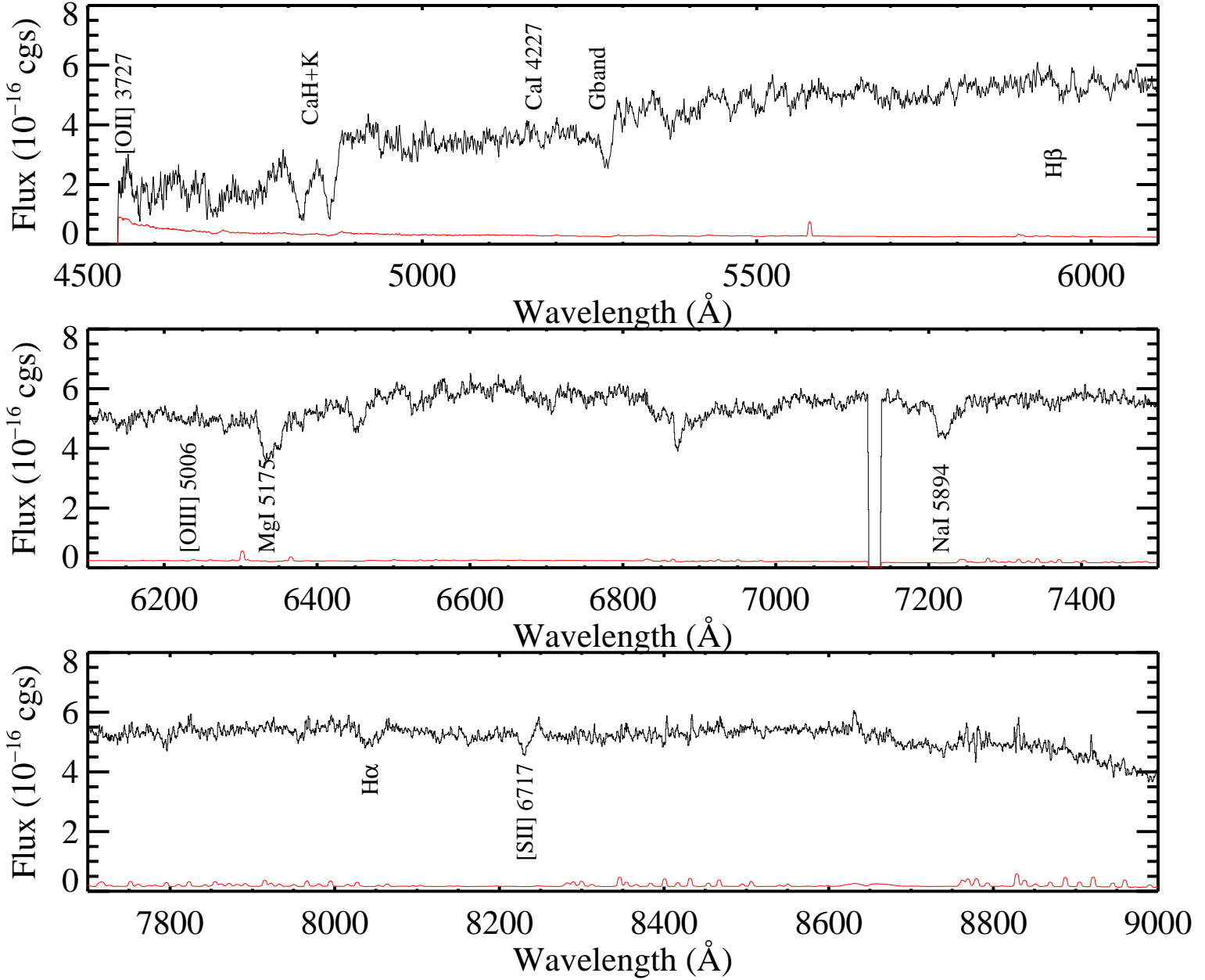}}
\caption[G1 Spectrum] {Keck/DEIMOS spectrum of the galaxy G1 (along
with its variance spectrum) located $\sim 10$\arcsec\ west of the
center of the XRT error circle for GRB\,050509b.  The data were
obtained using the 600 line mm$^{-1}$ grating centered at 7200~\AA\ and the
galaxy was observed through a 1.1\arcsec\ slit (FWHM $\approx 5$~\AA).
The strong absorption-line features indicate $z=0.2248$, and a
comparison of the spectrum against a template spectrum of HD~72324
provides an estimate of the velocity dispersion: $\sigma = 260 \pm 40$
\kms.}
\label{fig:galxspec}
\end{figure*}

Figure~\ref{fig:galxspec} presents the 1D flux-calibrated spectrum of
G1 against a vacuum, heliocentric-corrected wavelength array.  The
dotted line traces a 1$\sigma$ error array based on Poisson counting
statistics. We have marked a number of detected absorption-line
features and also the expected position for several strong transitions
frequently observed in emission-line galaxies (e.g., H$\alpha$, 
[O~III]).  We have fit a double-Gaussian profile to Ca~II H\&K and measure
$z=0.2248 \pm 0.0002$.  This is consistent with the redshift
inferred photometrically for this cluster from DPOSS \citep{gcl+03}.
At this redshift, the luminosity distance is 1117.4~Mpc, and 1\arcsec\
corresponds to 3.61~kpc in projection.

To estimate the velocity dispersion of the galaxy, we have compared
the spectrum against a template spectrum of HD~72324 (e.g.,
\citealt{kivf00}) smoothed by a wide range of $\sigma$.  The best
match to the absorption lines of G1 with $\lambda_{rest} =
4000$--5300~\AA\ is $\sigma=275 \pm 40$ \kms.  Accounting for the
instrumental resolution, we derive a light-weighted velocity
dispersion for this galaxy of $260 \pm 40$\kms.

The spectral features evident in Figure~\ref{fig:galxspec} are typical
of early-type galaxies.  The spectral type and velocity dispersion
indicate a massive elliptical galaxy with no apparent ongoing star
formation.  A quantitative limit to the current star-formation rate
(SFR) can be inferred from the upper limit to the H$\alpha$ luminosity
of this galaxy.  The emission-line flux in a 10~\AA\ window ($\Delta v
\approx 300$ km s$^{-1}$) centered at the expected wavelength of
H$\alpha$ has a 3$\sigma$ upper limit of $1.2 \times 10^{-16}$ erg
s$^{-1}$ cm$^{-2}$.  Adopting the current concordance cosmology, we
derive an H$\alpha$ luminosity $L_{H\alpha} < 1.2 \times 10^{40}$ erg
s$^{-1}$. Using the empirical relation between SFR and $L_{H\alpha}$
\citep{ken98}, the $3\sigma$ upper limit to the current SFR is
0.1~$M_\odot$ yr$^{-1}$.

A morphological fit to WIYN $I$-band imaging using {\it galfit}
\citep{phir02} shows good agreement with a de Vaucouleurs profile
(S\'ersic index = 4), with a $\chi^2$/dof = 1.22. The effective radius
is $R_e$ = 0.96\arcsec\ = 3.47~kpc. The galaxy has an axis ratio of
0.81 with the semimajor axis aligned along a position angle east of
north at $\sim$90\arcdeg. There was little improvement in $\chi^2$/dof
by adding more complicated morphologies or letting the S\'ersic index
vary.

The coincidence of a point source at radio wavelengths with the
optical center of G1 might suggest the presence of a low-level active
galactic nucleus despite the lack of telltale features observed in the G1 
optical spectra. Moreover, inspection of archival
images of this galaxy from the Near-Earth Asteroid Tracking Program
\citep{pph+99} on 9 April 2002, 20 April 2002, 3 May 2002, 22 March
2003, and 8 April 2003 reveals no apparent variability of the optical light
from G1. However, radio emission without corresponding optical emission is not
uncommon in giant elliptical galaxies harboring mildly active nuclei
\citep{ho99}.  The radio emission in G1 is unlikely to be associated with 
star formation, given the low SFR deduced above.

The properties of this probable host galaxy contrast significantly
with those measured for the galaxy hosts of long-duration GRBs.
First, most hosts of long-duration GRBs exhibit emission-line features
indicative of high SFRs \citepeg{dfk+03}. Second, the absolute
$K$-band luminosity of this galaxy ($\approx$1.6 $\times 10^{11}
L_\odot$) exceeds that of all previously identified GRB host
galaxies \citep{cba02}.  Third, the impact parameter of the GRB (as
defined by the 90\% XRT error circle) is larger than that of all
previously associated GRB-host galaxy pairs (long-burst offsets
$\ale$10 kpc; \citealt{bkd02}).

\subsection{S1, S2: Faint Blue Galaxies in a High-Redshift Group?}
\label{sec:smurfs}

The Gemini/GMOS spectra of S1 and S2 are featureless and blue.
Examining the regions of the spectrum where H$\alpha$ or H$\beta$
would lie if at the redshift of the cluster, we detect no measurable
emission.  Assuming, for the moment, that the sources are at the
cluster redshift of 0.22, we put a 3$\sigma$ upper limit on the
H$\alpha$ luminosity of $L_{H_{\alpha}} < 1.5 \times 10^{39}$ erg
s$^{-1}$ and $L_{H_{\alpha}} < 1.4 \times 10^{39}$ erg s$^{-1}$ for S1
and S2, respectively.  Using the equation from \citet{ken98} relating
the H$\alpha$ luminosity to the SFR, we find that the upper limits for
the unextinguished SFR, assuming that S1 and S2 are cluster members,
are $\sim 1.1 \times 10^{-2} M_{\sun}$\,yr$^{-1}$ for the galaxies.
If S1 and S2 are cluster members, then they are not forming stars,
which would seem to conflict with their blue colors.

A more likely scenario, also mentioned by \citet{csb+05b}, is that S1
and S2 are both background galaxies.  Although the Gemini spectra
range from 4600~\AA\ to 8600~\AA, the data have poor signal-to-noise
ratio blueward of 5200~\AA.  Nevertheless, the spectral slope is well
constrained and it suggests that these galaxies are forming stars
(i.e., the continuum is relatively blue).  The lack of corresponding
emission lines (H$\alpha$, H$\beta$, [O~III] $\lambda$5007, [O~II]
$\lambda\lambda$3727) falling in our spectral window therefore suggests that
S1 and S2 have $z \age 1.3$.  Additional spectroscopy will be required
to confirm our hypothesis that S1 and S2 are faint blue galaxy members
of a small group at moderate redshift.

\section{Theoretical Interpretation}
\label{sec:theory}

\begin{figure*}[p]
\centerline{\includegraphics[width=5.5in,angle=0]{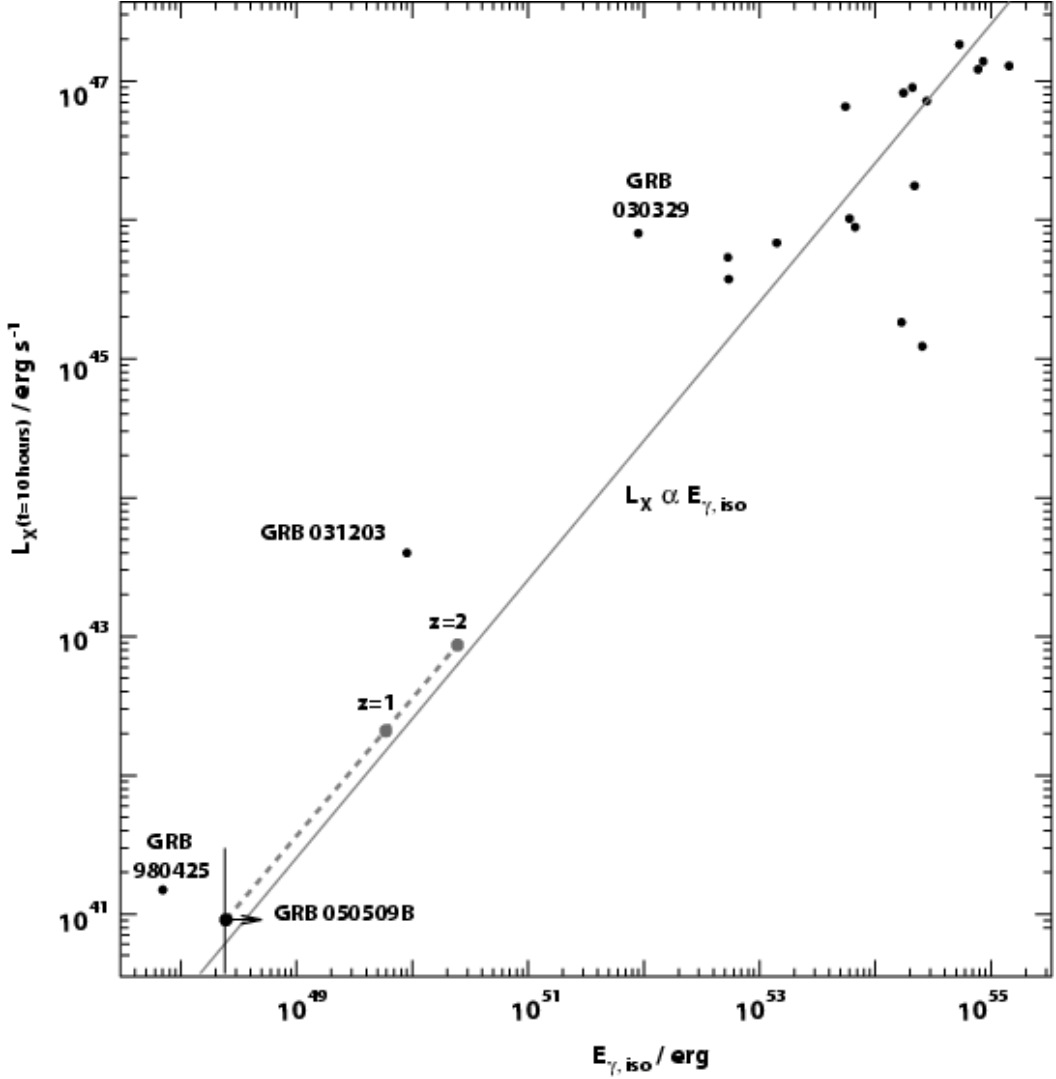}}
\caption{{\small Isotropic equivalent luminosity of GRB X-ray
afterglows scaled to $t = 10$ hr (source frame) after the burst as a
function of their isotropic gamma-ray energy release (adapted from
\citealt{kwp+04}).  If GRB\,050509b is located at $z=0.2248$, the
isotropic equivalent luminosity of the X-ray transient at $t = 10$~hr
assuming $t^{-1.3}$ and the isotropic gamma-ray energy would be $\sim
9 \times 10^{40}$ erg s$^{-1}$ and $\sim 2.7 \times 10^{48}$ erg,
respectively (black symbol).} }
\label{fig:lxeiso}
\end{figure*}

The fluence of the prompt gamma-ray emission measured by the {\it
Swift} BAT is $f=(2.3 \pm 0.9) \times 10^{-8} {\rm ~erg~cm^{-2}}$
\citep{barth0509b}, which at the redshift of the tentative host
implies an isotropic equivalent energy output of
$E_{\rm\gamma,iso}=(2.7 \pm 1)\times 10^{48}$ erg. Since $\nu F_\nu$
is still rising roughly as $\nu^{0.5}$ in the 15--150~keV {\it Swift}
range, the total fluence could be $\gtrsim 3$ times larger if the peak
energy $E_p\gtrsim 1$--2 MeV.  Figure \ref{fig:lxeiso} shows the
isotropic equivalent luminosity of GRB X-ray afterglows scaled to $t =
10$ hr after the burst (in the cosmological rest frame of the source),
$L_X(10\,{\rm hr})$, as a function of their isotropic gamma-ray energy
release, $E_{\rm\gamma,iso}$, for GRB\,050509b together with a sample
of long GRBs.  $L_X(10\,{\rm hr})$ for GRB\,050509b is estimated by
extrapolating the flux measured by the {\it Swift} XRT using the
best-fit power-law decay index of $\alpha=1.3$, which is also
consistent with the {\it Chandra} upper limit.

A linear relation, $L_X(10\,{\rm hr}) \propto E_{\rm\gamma,iso}$,
seems to be broadly consistent with the data, probably suggesting a
roughly universal efficiency for converting kinetic energy into
gamma rays in the prompt emission for both short and long
GRBs. This ``universal'' efficiency is also likely to be high (i.e.,
the remaining kinetic energy is comparable to, or even smaller than,
that which was dissipated and radiated in the prompt emission). If
this is the case, the well-known efficiency problem for long GRBs also
persists for short GRBs.

The X-ray luminosity at 10~hr is used as an approximate estimator for
the energy in the afterglow shock, since (a) at 10~hr the X-ray band
is typically above both $\nu_m$ and $\nu_c$ so that the flux has a very
weak dependence on $\epsilon_B$ [to the power of $(p-2)/4$] and no
dependence on the external density, both of which have relatively
large uncertainties \citep{fw01,pkpp01,bkf03}; and (b) at 10~hr the
Lorentz factor of the afterglow shock is sufficiently small ($\Gamma
\approx 10$) so that a large fraction of the jet is visible (out to an
angle of $\sim \Gamma^{-1} \approx 0.1$ rad around the line of sight)
and local inhomogeneities on small angular scales are averaged out.
Furthermore, the fact that the ratio of $L_X(10\,{\rm hr})$ and
$E_{\rm\gamma,iso}$ is fairly constant for most GRBs suggests that
both can serve as reasonable measures of the isotropic equivalent
energy content of the ejected outflow. A possible caveat to the above
statement arises if the observer is in fact not within the aperture of
the GRB jet (as is suggested to be the case in both X-ray flashes 
and X-ray rich GRBs;
\citealt{grp05}). In this case $E_{\rm\gamma,iso}$ can be
significantly smaller than the isotropic equivalent kinetic energy in
the afterglow shock, which is better reflected by $L_X(10\;{\rm
hr})$. This is likely to be the reason why GRB\,031203 is above the
correlation shown in Figure~\ref{fig:lxeiso} \citep{ram05}.
An off-axis interpretation for GRB\,050509b, on the other
hand, is unlikely since its X-ray afterglow light curve was observed
to decay from a very early epoch (at $t \approx 10^2$~s). This is also
consistent with the fact that GRB\,050509b falls close to the
correlation.

The above arguments suggest that the energy in the outflow ejected by
GRB\,050509b was $\sim E_{\rm\gamma,iso}\approx 10^{48.5}$\,erg, if it
was spherical. On the other hand, if it was collimated into a narrow jet of 
half-opening angle $\theta_0$, then the true energy would be smaller by a
factor of $f_b=(1-\cos\theta_0) \approx \theta_0^2/2$. Since a
significant off-axis viewing angle is not likely, the true energy
probably does not exceed $E_{\rm\gamma,iso}$. A higher redshift would
increase $E_{\rm\gamma,iso}$ and with it the estimate for the energy
release in this event; however, it would still remain significantly
less energetic than typical long GRBs (see Figure~\ref{fig:lxeiso}).

\begin{figure*}[p]
\centerline{\includegraphics[width=5in,angle=0]{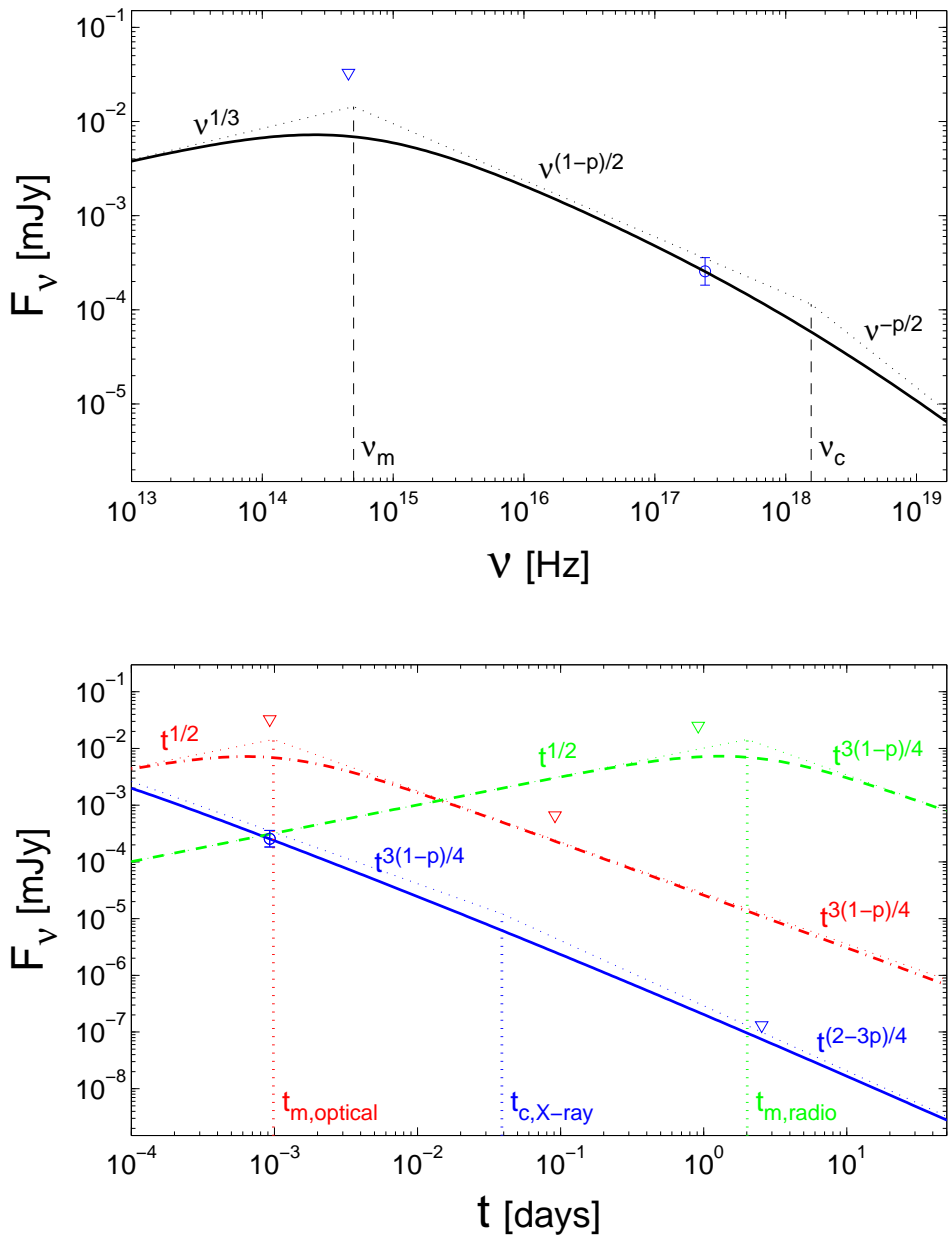}}
\caption{{\small The spectrum at $t = 80$~s after the trigger ({\it
upper panel}) and the light curves ({\it lower panel}) of the X-ray
(1~keV; blue solid line), optical ($R$ band; red dashed-dotted line),
and radio (4.9~GHz; green dashed line) emission for a spherical
afterglow shock propagating into a uniform ambient medium, using the
model of \citet{gs02}. We also show the flux normalization in the
X-rays from the {\it Swift} XRT detection, as well as upper limits in
the optical and in the radio. Here, we adopt the redshift of the
tentative host galaxy ($z=0.2248$) with {\it typical} interstellar
medium density $n=1$~cm$^{-3}$, and assume that the isotropic
equivalent kinetic energy in the afterglow shock is equal to
$E_{\gamma,{\rm iso}}$ (i.e., $2.7 \times 10^{48}$ erg). The
microphysical parameters are taken to be typical of those inferred
from the modeling of afterglows of long GRBs: $p=2.2$,
$\epsilon_E=0.15$, $\epsilon_B=0.046$. Table \ref{tab:theory} gives
other models consistent with the data.}}
\label{fig:speclc}
\end{figure*}

As also argued by \citet{lrrg05}, the fact that the X-ray afterglow
luminosity of GRB\,050509b is much smaller than that of long GRBs is
probably because the event was sub-energetic, rather than due to
differences on the values of the external density or the microphysical
parameters. This is illustrated in Figure~\ref{fig:speclc} by a fit to
the currently available afterglow data using parameter values that are
typical for long GRBs, except for the isotropic equivalent energy in
the afterglow shock, $E_{\rm k,iso}$, which is here taken to be equal
to $E_{\rm\gamma,iso}$ assuming $z=0.2248$. Other parameter values
could also give a reasonable description of the rather sparse data.
In Table~\ref{tab:theory} we demonstrate a few different sets of
parameters that fit the afterglow data.  Again, we refer the reader to
\citet{lrrg05} for a more detailed description. Regardless of the
redshift, it will be very difficult to detect the afterglow in the
radio, since the maximal flux density (given the observational
constraints) is unlikely to exceed $\sim 15~\mu$Jy.

If short GRBs occur significantly outside of their host galaxies, as
may be common for binary mergers \citep{ty94,bsp99,fwh99,bbz99,bbz00},
then one might expect the external density encountered by the
afterglow shock of some GRBs to be very low, typical of the
intergalactic medium (IGM), $n_{\rm IGM} \approx
10^{-6.5}(1+z)^3\;{\rm cm^{-3}}$. This may help explain why some short
bursts could have very faint afterglows. Since GRB\,050509b happened
to occur near the center of a galaxy cluster where the external
density is relatively high, its X-ray afterglow was relatively
brighter. If indeed GRB\,050509b is associated with the galaxy cluster
at $z\approx 0.22$, then one might expect the external density to be
intermediate between the IGM and ISM: the GRB is $\sim76$\arcsec\ from
the center of the cluster as determined by the X-ray position
(\S~3.4), corresponding to $\sim 270$ kpc in projection and well
within the diffuse emission from the hot intracluster medium gas
(which extends to a radius of $\sim1$~Mpc). This suggests an ambient
density near the position of the GRB of $n \approx 10^{-3}$--$10^{-2}$
cm$^{-3}$, though this estimate is uncertain because the space
position of the burst relative to the cluster center and the
intracluster medium (ICM) density profile are not known precisely.

\section{Discussion}

The lack of a strong afterglow signature sets GRB\,050509b apart from
most other GRBs.\footnotemark\footnotetext{We note that two other
low-luminosity, low-redshift GRBs (980425 and 031203) had detectable
X-ray afterglow emission but no optical afterglow. The only transient
optical signatures for these bursts were apparent supernovae.} As a
comparison, the low-redshift long-soft burst (GRB\,030329, $z=0.1685$;
\citealt{gkr+03}), if placed at the redshift of G1, would have been $R
\approx 14$\,mag at $t=8000$~s; this is approximate 10 mag brighter
than the detection limits found herein. Even at $z=3$, the optical
afterglow of a GRB\,030329-like burst should have been detected at
early times (neglecting the effects of dust extinction). Our
non-detections ($R \age 24$\, mag) of variability at 1.3 hr in the
what would be the restframe at $z$=1 is more than 3.5 magnitudes
deeper than the faintest optical tranisent found for a long GRB
(021211; $z$=1.0; see fig.\ 2 of \citealt{fps+03}).

The lack of detectable optical/infrared afterglow is not surprising on grounds
related to the progenitors and to GRB afterglow theory. First, since the
luminosity of long-wavelength afterglows scales with the square root
of the ambient density \citep{bmr93,mrw98b}, events that occur in the
ISM or IGM should be intrinsically fainter (at optical/infrared wavelengths)
than those occurring in the circumburst environments of collapsars
(see \citealt{pkn01}).  Second, based on $<V/V_{\rm max}>$ studies,
the isotropic-equivalent peak luminosity ($L_p(\gamma)$) of short bursts
is similar to that of long bursts \citep{sch01}, implying that the
total energy output ($E_{\rm\gamma,iso} \approx L_p(\gamma)/\eta
\times$ duration, with $\eta$ as the conversion efficiency to 
gamma rays) is at least an order of magnitude smaller for short bursts. 
As argued by \citet{pkn01}, since afterglow brightness scales with
$E_{\rm\gamma,iso} (1 - \eta)$, short-burst afterglows would be
systematically faint.

Now that there is a detected X-ray afterglow we are in a position to
directly test the faintness claim, by inferring the gamma-ray
energy release and X-ray afterglow luminosity (a proxy for the kinetic
energy in the blast wave). From Figure \ref{fig:lxeiso} it is clear
that this ratio for GRB\,050509b is similar to that found in
long-duration GRBs. This is a striking observational bridge to
long-duration bursts and suggests a common physical mechanisms for
prompt and delayed (afterglow) emission for both long-duration and
short-duration GRBs, even though their progenitors are probably
different.

A tentative detection of an afterglow signal by adding up the emission
of 76 short BATSE bursts was reported by \citet{lrg01} (see also
\citealt{conn02}). The signal peaked at $t\approx 30\;$s after the
burst trigger with a relatively flat $\nu F_\nu \approx 5\times
10^{-10}\;{\rm erg\;cm^{-2}\;s^{-1}}$. This would correspond to an
X-ray flux in the 0.2--10~keV range, of $F_X \approx 2\times
10^{-9}\;{\rm erg\;cm^{-2}\;s^{-1}}$. The X-ray flux of the afterglow
of GRB\,050509b is best constrained around $t\approx 200$~s, and is 
found to be $F_X\approx6.5\times 10^{-13}\;{\rm
erg\;cm^{-2}\;s^{-1}}$. Extrapolating this flux to $t\approx 30$~s
with a power-law index in the range inferred from the data,
$1.0\lesssim\alpha\lesssim 1.7$, gives a flux that is lower than the
one found by \citeauthor{lrg01} by a factor of $\sim(1-5)\times
10^2$. This might suggest that either the possible detection by
\citet{lrg01} was not statistically significant, or the X-ray
afterglow of GRB\,050509b is underluminous compared to the average
value for short GRBs by at least two orders of magnitude.

With essentially no indication of recent star formation in G1,
massive progenitor stars leading to collapsars cannot be present in
G1.  S1 and S2, the brightest and third-brightest sources within the
XRT error circle, have no indication of recent star formation if their
redshifts are $\lesssim 1.3$ (SFR $< 0.05 M_{\sun}$ year$^{-1}$ for $z
< 0.3$ and SFR $< 1 M_{\sun}$ year$^{-1}$ for $z < 1.2$).  The fainter
(and blue) objects discussed in \S~\ref{sec:smurfs} are likely to
be background galaxies.  If the origin of GRB\,050509b is from a
collapsar, it is likely that its redshift exceeds 1.3.

If GRB\,050509b is a background object at $z \gtrsim 2$, some
progenitor scenarios are difficult to reconcile.  With an observed
duration of $\sim 30$ msec, the rest-frame duration would be only
about $10$ msec.  This is implausibly short for an NS--NS merger, and
marginally possible for a BH--NS merger if the coalescence is through
unstable mass transfer \citep{lee99,ros05,mcm05}.  It is hard to
simultaneously accommodate the short intrinsic timescale and the
higher energy budget of the burst within any compact merger model, if
it is at high redshift.

If short GRBs trace star formation with a time delay through double
compact mergers with coalescence time scales of $10^{7}$--$10^{10}$ yr
(as opposed to prompt tracers of star formation as with the collapsar
scenario for long GRBs; \citealt{bpzy98} and \citealt{bsp99}), then we
expect some fraction (10--30\%) of short GRBs to be seen in association
with early-type galaxies in general and clusters specifically (see
\citealt{nut04} for rate density in the local universe).  This is
somewhat model dependent, since the distribution of compact merger
timescales is poorly constrained by data, but broadly consistent with
both observed and model distributions.

A core-collapse supernova (SN) produces no electromagnetic radiation
until its envelope is completely consumed by the explosion (although
see \citealt{kho99}). This phase ends, however, with a brilliant flash
of X-ray or extreme ultraviolet photons as the shock reaches the
stellar surface. The ``breakout'' flash is delayed in time, and vastly
reduced in energy, relative to the neutrino transient produced by core
collapse. However, it conveys useful information about the
explosion. Shock breakout flashes were predicted by \citet{col68} as a
source for (the then undetected) gamma-ray bursts. The explosion of SN
1987A stimulated a reanalysis of supernova breakout flashes by
\citet{eb92} and, more recently, by \citet{beb+98} and
\citet{blb+00}. These studies represent an increase in sophistication
toward the full numerical treatment of this complicated,
radiation-hydrodynamic problem.  In principle, the XRT data could
constrain the existence of a shock breakout produced by both a red
supergiant explosion like SN~1993J \citep{van02} and a blue supergiant
explosion analog to SN~1987A, but the X-ray luminosity is sensitive to
the uncertain distribution of extragalactic gas column and the
specific XRT observing epochs.

Using our ESI optical imaging, we can also limit the presence of
brightening due to a supernova or supernova-like emission at 8.17~d
after the GRB to $R_c \approx 25.0$ mag. A normal, unextinguished Type
Ia (thermonuclear) supernova at $z = 0.22$ would have $R \approx 22$
mag, around 6.7~d after explosion ($t = 8.17$~d in the observer's
frame). A very subluminous SN~Ia like SN 1991bg \citep{frb+92} would
have $R \approx 24$ mag, still somewhat brighter than our
limit. Extinction would obviously make the SN fainter, but the Milky
Way contribution is small ($A_V \approx 0.06$ mag; \citealt{sfd98}),
and the outskirts of an elliptical galaxy in a cluster should have
essentially no dust.  While some core-collapse supernovae could be as
faint as (or fainter than) our limit, the presence of such a supernova
in the outskirts of an elliptical galaxy would be truly extraordinary
(see \citealt{vwf05}). Others have also reported no evidence for a SN
at later-times \citep{hsg+05,bfr+05}.

The location of this (and future) short burst provides a useful
discriminant for distinguishing between different progenitor models of
short bursts. Simplistically, we would expect evaporating black holes
to occur near the center of deep potential wells (as discussed in the
context of Galactic BHs; \citealt{cmo99}); thus, the offset from G1
seems to disfavor this hypothesis.  A giant flare from a magnetar
would need to have a isotropic luminosity ($L_{\rm\gamma,iso}$) larger
by a factor of $\sim 10^3$ and an $E_{\rm\gamma,iso}$ larger by a
factor of $\sim 10^2$ compared the the initial spike of the 27
December 2004 giant flare from SGR 1806-20 (the difference in the
factor between the two quantities arises since GRB\,050509b lasted
only $\sim 30$~ms, which is $\sim 10$ times shorter than the initial
spike of the giant flare from SGR 1806-20). Bursts from magnetars
might be expected from later-type galaxies than G1 where neutron stars
would be formed copiously: magnetic field decay would cut the active
lifetime for megaflare activity after $\sim10^4$ yr.

\section{Conclusions}

We have monitored the location of GRB\,050509b at optical and infrared
wavelengths from 8 minutes to 8 days after the trigger and found no
indication of variability at the location of the fading X-ray source,
the first solid X-ray detection of an afterglow of a short-hard burst.
Near the location of this source we and others have found an apparent
group of faint blue galaxies at redshifts $\gtrsim 1.3$. While it is
indeed plausible that this short burst arose from a progenitor
connected with those galaxies, we found --- based on a positional
argument --- plausible evidence that the progenitor is likely
associated with G1, a bright elliptical galaxy at $z=0.2248$.  We have
argued that the observations find natural explanation with a compact
merger system progenitor. If so, then short-hard GRBs provide a bridge
from electromagnetic to gravitational wave astronomy: indeed has
GRB\,050509b occurred a factor of $\sim$3 closer in luminosity
distance is might have produced a detectable chirp signal with the
next generation LIGO-II gravitational wave
facility\footnotemark\footnotetext{\url{http://www.ligo.caltech.edu/docs/G/G990111-00.pdf}}.

Brightening emission from most types of supernovae would have been
seen in our imaging, so the lack of such emission appears inconsistent
with the notion that short bursts are due to collapsars or variants
thereof. Our afterglow modeling is also consistent with, though does
not require, a circumburst medium having lower density than that
inferred in long-duration GRBs; if true, this would suggest that the
progenitor produces a GRB in an environment that is baryon poor
compared to that expected for collapsars.  Moreover, we have seen no
evidence for ongoing star formation in the putative host, so there are
likely no remaining massive stars. Given the short active life of a
neutron star having a high magnetic field, this also disfavors the
magnetar hypothesis.

\begin{figure*}[p] 
\centerline{\includegraphics[width=5.5in,angle=0]{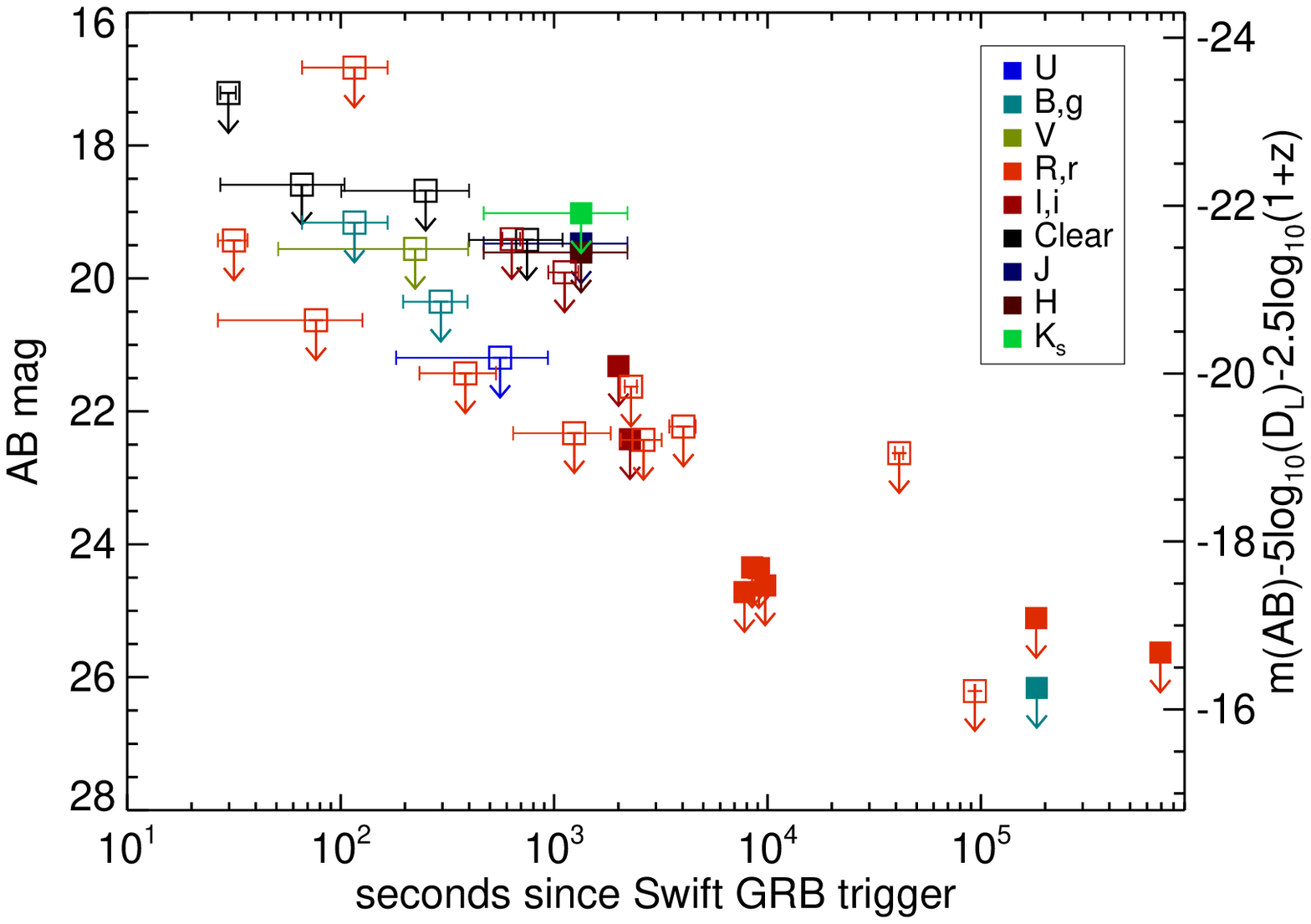}}
\caption[Upper limit light curve] {\small Upper limits from WIYN,
  PAIRITEL, and Keck images (filled symbols) along with those reported
  in the literature (open symbols). Magnitudes are corrected for
  extinction using the maps of \citet{sfd98} and converted to AB
  magnitudes, assuming $z=0.2248$ (see text), using the relations of
  \citet{fg94} and \citet{bbc+03}. Times are reported in seconds
  relative to the {\it Swift} trigger on May 9.166889. The WIYN and
  PAIRITEL upper limits are $5\sigma$, but many of the quoted limits
  in the literature are not accompanied by a stated significance
  level. For smaller telescopes with large pixels, the light from the
  nearby galaxy is likely a significant contaminant, resulting in
  upper limits that may be overestimated in the literature. }
\label{fig:upper}
\end{figure*}

The non-detection of brightening emission may place limits on the
presence of a thermal ``mini-supernova'' from non-relativistic ejecta
of a compact merger system \citep{lp98b,ros02}. In this scenario, the
small dense mass ($m_{\rm ej}$) ejected during coalescence expands as
it is heated by radioactivity of the decompressed ejecta. Using the
scalings of \citet{lp98b} and crudely assuming that 10\% of the
bolometric light at peak is radiated in the $R$ band, the $R$-band
brightness should peak at observer time $t \approx 1.2 ({m_{\rm
ej}/0.01 M_\odot})^{1/2}$~d after the burst, with absolute magnitude
$M_R \approx -18.5 - 1.25 \log ({m_{\rm ej}/0.01 M_\odot})$
mag. Assuming that the GRB did indeed originate from the redshift
$z=0.2248$, upon inspection of Figure \ref{fig:upper}, with
non-detections at $M_R \approx -16$ mag at $t \approx 1$~d, we can
very roughly exclude $m_{\rm ej} > $few $\times$10$^{-3}
M_\odot$. Though the \citet{lp98b} model was intended as a simplistic
sketch of the phenomenon, this limit on $m_{\rm ej}$ is somewhat
surprising given the amount of escaping non-relativistic material
expected in compact mergers \citep{ros02}. Indeed, we consider this
lack of a ``mini-supernova'' as weak evidence against a $z=0.22$
origin from a compact merger system. Still, these limits are subject to
considerable uncertainty in a number of uncertain parameters of
ejecta. For instance, if the velocity of the ejecta were to be
$\sim$0.01$c$ instead of 0.3$c$ (as assumed by \citealt{lp98b}) then
the peak of the thermal emission would occur after about 1 month, and
would not have been detected with the current limits.

We conclude by emphasizing that in the NS--NS or BH--NS progenitor
hypothesis for short-hard bursts, the hosts galaxies may be a range of
Hubble types \citepeg{lsp+98}. Compact merger systems coalesce in
appreciable rates from Myr to Gyr after a starburst
\citepeg{fwh99,bsp99}. Obviously, the longer the time since the
starburst, the larger the distance a binary system will travel before
coalescence. A clear prediction from this model is that as more short
bursts are localized, those associated with later-type galaxies of
{\it a given mass} should be preferentially closer to the
star-formation centers of the host; that is, we expect a more
concentrated distribution around a spiral galaxy with the same mass as
an early-type. On the other hand, dwarf star-forming hosts have
shallow enough potentials that merger systems from these galaxies
could coalesce at appreciable distances ($\age 100$kpc) even shortly
after starburst. As {\it Swift} localizes more short-hard bursts, we
expect that the offset distribution around galaxies will further
elucidate the progenitor question.

\acknowledgements

We thank the anonymous referee for their insightful and detailed
comments that lead to the improvement of this paper. Much of this work
could not have been undertaken without the work of S. Barthelmy
running the GCN Circulars and the {\it Swift} team having deployed
such a marvelous experiment. We thank the Gemini staff for their
excellent work in observing GRB\,050509b, especially K. Roth. We thank
R.~Duncan, A.~Loeb, P.~Kumar, and E.~Quataert for informative
exchanges. We thank B.~Simmons for her assistance with using {\it
galfit}. We also thank M. Skrutskie, D. Starr, W. Peters, B. Hutchins
for their contributions to and assistance with PAIRITEL. We are
indebted to G. Wirth and R. Chornock for their contributions in
acquiring some of the Keck data. J.S.B., J.X.P., and H.-W.C. are
partially supported by NASA/{\it Swift} grant NNG05GF55G. The work of
A.V.F. is supported by NSF grant AST-0307894 and NASA/{\it Swift}
grant NNG05GF35G; he is also grateful for a Miller Research
Professorship at U.C. Berkeley, during which part of this work was
completed. S.S. is supported by the Center for Gravitational Wave
Physics funded by the NSF under cooperative agreement PHY 01-14375,
and by NSF grant PHY 02-03046. D.P. and E. R.-R. gratefully
acknowledge support provided by NASA through Chandra Postdoctoral
Fellowship grant numbers PF4-50035 and PF3-40028 (respectively),
awarded by the Chandra X-ray Center, which is operated by the
Smithsonian Astrophysical Observatory for NASA under contract
NAS8-03060. S.J. gratefully acknowledges support via a Miller Research
Fellowship at U.C. Berkeley.  J.N. acknowledges support from NASA
through Hubble Fellowship grant HST-HF-01165.01 awarded by the Space
Telescope Science Institute, which is operated by the Association of
Universities for Research in Astronomy, Inc., under contract NAS
5-26555.  The research of J.G. is supported by the US Department of
Energy under contract number DE-AC03-76SF00515.  This publication
makes use of data products from the Two Micron All Sky Survey, which
is a joint project of the University of Massachusetts and the Infrared
Processing and Analysis Center/California Institute of Technology,
funded by the National Aeronautics and Space Administration and the
National Science Foundation. Some of the data presented herein were
obtained at the W. M. Keck Observatory, which is operated as a
scientific partnership among the California Institute of Technology,
the University of California, and NASA; the Observatory was made
possible by the generous financial support of the W. M. Keck
Foundation.  The analysis pipeline used to reduce the DEIMOS data was
developed at U.C. Berkeley with support from NSF grant AST-0071048.
Based on observations obtained at the Gemini Observatory, which is
operated by the Association of Universities for Research in Astronomy,
Inc., under a cooperative agreement with the NSF on behalf of the
Gemini partnership: the National Science Foundation (United States),
the Particle Physics and Astronomy Research Council (United Kingdom),
the National Research Council (Canada), CONICYT (Chile), the
Australian Research Council (Australia), CNPq (Brazil) and CONICET
(Argentina).  Finally, we wish to extend special thanks to those of
Hawaiian ancestry on whose sacred mountain we are privileged to be
guests.

%\bibliographystyle{apj}
%\bibliography{journals_apj,grbrefs,grbrefs-unpublished,gcns}

\clearpage

\begin{figure*}[p] 
\centerline{\includegraphics[width=5.in,angle=270]{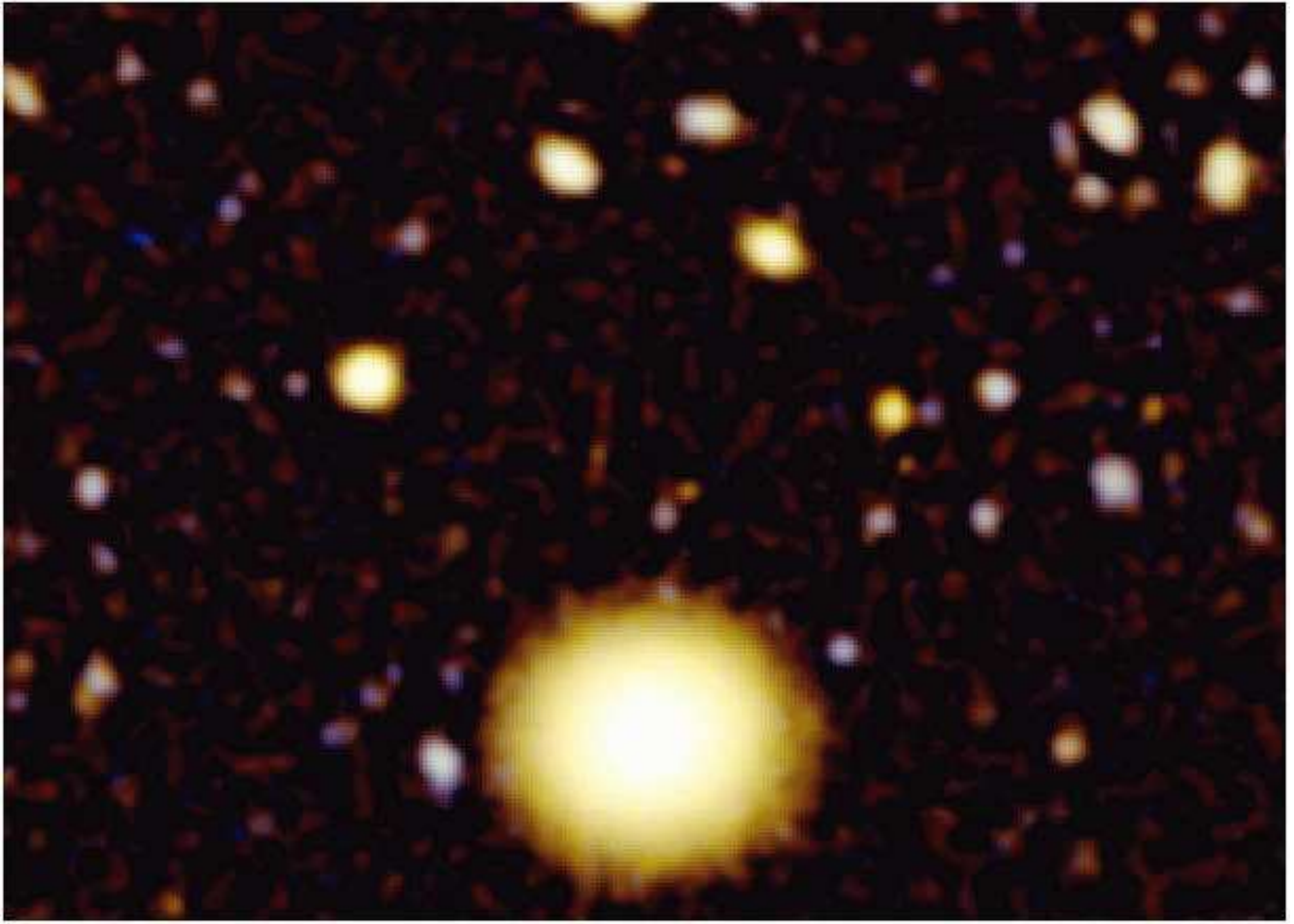}}
\caption[XRT Error Circle] {False-color image of the field of XRT
constructed with the $G$ (blue) and $R$ (red) Keck/LRIS images; green is
interpolated between the observed bands. Aside from source S6 (which
appears red) and possibly the J2/J4 complex, all of the
XRT-consistent sources appear to be faint and blue, consistent with a
small group of star-forming galaxies at a redshift larger than the
cluster. As seen in this image, a number of such groups appear
throughout the field (there are 2 faint blue galaxies to the north of
G1, also embedded in the light of G1).}
\label{fig:lris-col}
\end{figure*} 

\begin{deluxetable}{lcccccr}
\singlespace
%\rotate
\tablecolumns{7}
\tablewidth{0in}
\tablecaption{Upper Limits on Optical-IR Afterglow of GRB\,050509b\label{tab:upper}}
\tabletypesize{\small}
\tablehead{
\colhead{Start Time\tablenotemark{a}} & \colhead{Exposure Time} & 
\colhead{Band} & \colhead{Limit\tablenotemark{b}}
& \colhead{Ref.} \\
\colhead{(s)} & \colhead{(s)} & \colhead{} &
\colhead{(AB mag)} & \colhead{}}
\startdata
   467.8   &        1741  &  $J$ &  19.3   & this work   \\
  467.8    &       1741   & $H$  & 19.5    & this work    \\
  467.8    &       1741   & $K_s$  &  18.95   & this work \\
1973.6    &  	60  &	$i$ &   20.95   &   this work \\
2118.3    &   	300 &	$i$ &   22.05   &   this work \\
7497.3    &     600 &	$r$     &  24.21&      this work             \\
8172.3    &   	600 &   $r$     &  23.84&      this work             \\
8808.3    &   	600 &	$r$     &  23.85&      this work             \\
9446.9    &   	600 &	$r$     &  24.11&      this work	           \\
179696    & 	1260 &  $R$	&  24.6 & 	this work \\
179696	  & 	1260 &  $G$	&  25.5	&  this work \\
691200	  &     960  &  $R$	&  25.1 & this work \\
27.3      &   	5   &   clear &  17.21&	\citet{rssq05}            \\
27.3      &   	77  &   clear &  18.59&	\citet{rssq05} 	    \\
100.6     &   	299 &	clear &  18.68&	\citet{rssq05}	    \\
399.1	  &   	696 &	clear &  19.42&	\citet{rssq05}	    \\
26.6	  &	10  &	$R$     &  18.92&	\citet{wvw+05}           \\
26.6	  &	100 &	$R$     &  20.12&   	\citet{wvw+05}   \\
234	  &	300 &	$R$     &  20.92&	\citet{wvw+05}	    \\
643	  &	1200&	$R$    &  21.82&	\citet{wvw+05}	    \\
2052	  &	1140&	$R$    &  21.92&	\citet{wvw+05}           \\
3466	  &	1140&	$R$    &  21.72& 	\citet{wvw+05}   \\
66	  &	100 &	$B$    &  18.34&	\citet{smna+05}	    \\	
66	  &     100  &	$R$    &   16.32 &  	\citet{smna+05}    \\
2142.9	  &   	300  &	$R$    &   21.12 &      \citet{uce+05}      \\
572.9	  &   	120  &	$I$    &   19.04	&\citet{tori05}		    \\
939.9     &   	360  &	$I$    &   19.54	&\citet{tori05}		    \\
51	  &   	345  &	$V$    &   18.96 &      \citet{brhr+05}    \\
196	  &   	197  &	$B$    &   20.14 &  	\citet{brhr+05}  \\
182	  &  	753  &	$U$   &   19.4	&\citet{brhr+05} \\
39600     &    3600  &  $R$    &  22.12  & \citet{mp05} \\
93240.    &     900  &  $R$     &  25.7   &     \citet{csb+05a}

\enddata

\tablenotetext{a}{Time since 04:00:19 UT, the time of the {\it
Swift}/BAT trigger.}

\tablenotetext{b}{Limits were converted to AB magnitudes following
\citep{fg94}. Limits are presumed to be 5$\sigma$ detection limits for
a point source (not all reports in the literature stated the
significance of the upper limits). No extinction correction has been applied to these
magnitudes.}

\end{deluxetable}

\begin{deluxetable}{lccccl}
\singlespace
%\rotate
\tablecolumns{7}
\tablewidth{0in}
\tablecaption{Properties of Faint Sources in the XRT Error Circle\label{tab:blobs}}
\tabletypesize{\small}
\tablehead{
\colhead{Source} & \multicolumn{2}{c}{Position\tablenotemark{a}} 
& \multicolumn{2}{c}{Magnitudes\tablenotemark{b}} & \colhead{Comments} \\
\colhead{} & \colhead{R.A.} & \colhead{Decl.} & \colhead{$g^\prime$} &
\colhead{$r^\prime$} & \colhead{}}
\startdata
S1 & 12:36:13.677 & +28:58:57.51 & $23.80 \pm 0.12$ & $22.80 \pm 0.14$
& marginally consistent \\
   &              &              &                  &                 
& with new XRT error circle \\
S2 & 12:36:14.100 & +28:58:58.90 & $25.27 \pm 0.15$ & $24.59 \pm 0.24$
& outside new XRT\\
   &              &              &                  &                 
& error circle \\
S3 & 12:36:13.961 & +28:59:00.87 & $25.79 \pm 0.21$ & $24.45 \pm 0.21$ & \\
S4 & 12:36:13.642 & +28:59:01.80 & $25.59 \pm 0.17$ & $25.52 \pm 0.47$ & \\
S5 & 12:36:13.865 & +28:59:02.05 & $25.76 \pm 0.20$ & $25.36 \pm 0.39$ & \\
S6 & 12:36:13.629 & +28:59:08.81 & $25.96 \pm 0.19$ & $24.46 \pm 0.22$ &  red
galaxy \\
J1 & 12:36:13.574 & +28:59:03.84 & $\sim$27.1 & $\sim$25.1 & \\
J2 & 12:36:13.464 & +28:59:05.18 & $\sim$28   & $\age$25.5    & 
marginal detection \\
J3 & 12:36:13.435 & +28:58:56.88 & $26.14 \pm 0.21$ & $24.27 \pm 0.29$ 
& outside new XRT\\
   &              &              &                  &                 
& error circle \\
J4 & 12:36:13.471 & +28:59:06.87 & $\sim$27.3 &  $\age$25.5 & 
marginal detection\\
   &              &              &         &         & 
related to J2? \\
J5 & 12:36:14.237 & +28:59:05.39 & $\sim$26.8 & $25.5 \pm 0.4$ 
& outside new XRT\\
   &              &              &                  &                 
& error circle \\
\\
\enddata

\tablenotetext{a}{J2000; Positions are based on image centroids and
have been tied to the SDSS astrometric frame (see text). The absolute
positional uncertainty, inherited in part from the SDSS uncertainties,
is $0.2$\arcsec. The relative uncertainty between object positions
is expected to be $\sim$0.1\arcsec.}

\tablenotetext{b}{Photometry has been tied to the SDSS calibrations of
the field. We used a color term in the conversion of the Keck $R$ to
the Sloan $r'$ filter and constant magnitude offset (following
\citealt{wbm+91}). We assumed no color term in converting $G$ to
$g^\prime$. Errors include the $0.1$ mag zero-point uncertainty in the
tie to the SDSS photometry. We do not report photometric errors for
those sources near the detection limit, but such errors are likely to
be of order 50\%. No extinction correction has been applied to these
magnitudes.}

\end{deluxetable}

\begin{deluxetable}{lllllll}
\tablecaption{{Representative Fits to the Afterglow of GRB~050509b\label{tab:theory}}}
\tablecolumns{7}
\tablewidth{4in}
\tablehead{
\colhead{$z$}& \colhead{$E_{k,\mathrm{iso}}$/erg}& 
\colhead{$n/{\rm cm^{-3}}$}&  \colhead{$\epsilon_e$}& 
\colhead{$\epsilon_B$}& \colhead{$p$}&  
\colhead{$E_{k,\mathrm{iso}}/E_{\gamma,\mathrm{iso}}$}}     
\startdata
 0.2248 & 2.75$\times 10^{48}$ &  1  &    0.15 &     0.046 &  2.2 &         1\\                   
0.2248  & 1$\times 10^{51}$ &   1$\times 10^{-6}$&   0.1 &      0.016 &  2.3     &    363   \\               
3   &   4.5$\times 10^{50}$  &  1 &     0.1  &     0.01 &   2.2  &       0.98 \\                
3    &  5.63$\times 10^{52}$ & 1$\times 10^{-6}$  &  0.1 &      0.01 &   2.3 &        122  \\                
\enddata
\tablecomments{An example of four different sets of parameters that fit the rather
sparse afterglow data for GRB 050509b. In the first two cases we fix the
redshift at that of the putative host galaxy (G1; $z=0.2248$), and assume an
external density that is either typical of the interstellar medium (ISM),
$1\;{\rm cm^{-3}}$, or typical of the intergalactic medium (IGM),
$10^{-6}\;{\rm cm^{-3}}$. In the last two cases we explore the option of a
relatively high redshift, $z=3$ with the same two very different values
for the external density. In all cases, the values of the micro-physical
parameters ($p$, $\epsilon_e$ and $\epsilon_B$) were chosen to be typical
of those inferred from afterglow fits for long GRBs.}
\end{deluxetable}

\end{document}